\documentclass[format=acmsmall, review=false, screen=true]{acmart}

\usepackage{tcolorbox}
\usepackage{multirow}
\usepackage{bm}
\usepackage[justification=centering]{caption}
\usepackage{balance}
\usepackage{float}
\usepackage{tabularx}
\usepackage{longtable}
\usepackage{makecell}
\usepackage{minted}
\usepackage[title]{appendix}
\usepackage{threeparttable}
\usepackage{subcaption}
\usepackage{listings}
\usepackage{xcolor}

\AtBeginDocument{%
  }

\setcopyright{acmlicensed}
\copyrightyear{2026}
\acmYear{2026}
\acmDOI{XXXXXXX.XXXXXXX}

\acmJournal{TOSEM}
\acmVolume{0}
\acmNumber{0}
\acmArticle{0}
\acmMonth{0}

\begin{document}

\title{A Comprehensive Evaluation of Parameter-Efficient Fine-Tuning on Code Smell Detection}

\author{Beiqi Zhang}
\email{zhangbeiqi@whu.edu.cn}
\orcid{0000-0003-1259-4312}
\affiliation{%
  \institution{School of Computer Science, Wuhan University}
  \city{Wuhan}
  \country{China}
}

\author{Peng Liang}
\email{liangp@whu.edu.cn}
\orcid{0000-0002-2056-5346}
\authornote{Corresponding author}
\affiliation{%
  \institution{School of Computer Science, Wuhan University}
  \city{Wuhan}
  \country{China}
}

\author{Xin Zhou}
\email{xinzhou.2020@phdcs.smu.edu.sg}
\orcid{0000-0002-4558-0622}
\affiliation{%
  \institution{School of Computing and Information Systems, Singapore Management University}
  \country{Singapore}
}

\author{Xiyu Zhou}
\email{xiyuzhou@whu.edu.cn}
\orcid{0009-0002-5946-0039}
\affiliation{%
  \institution{School of Computer Science, Wuhan University}
  \city{Wuhan}
  \country{China}
}

\author{David Lo}
\email{davidlo@smu.edu.sg}
\orcid{0000-0002-4367-7201}
\affiliation{%
  \institution{School of Computing and Information Systems, Singapore Management University}
  \country{Singapore}
}

\author{Qiong Feng}
\email{qiongfeng@njust.edu.cn}
\orcid{0000-0003-1667-8062}
\affiliation{%
  \institution{School of Computer Science, Nanjing University of Science and Technology}
  \city{Nanjing}
  \country{China}
}

\author{Zengyang Li}
\email{zengyangli@ccnu.edu.cn}
\orcid{0000-0002-7258-993X}
\affiliation{%
  \institution{School of Computer Science, Central China Normal University}
  \city{Wuhan}
  \country{China}
}

\author{Lin Li}
\email{cathylilin@whut.edu.cn}
\orcid{0000-0001-7553-6916}
\affiliation{%
  \institution{School of Computer Science and Artificial Intelligence, Wuhan University of Technology}
  \city{Wuhan}
  \country{China}
}

\renewcommand{\shortauthors}{Zhang et al.}

\begin{abstract}
\textcolor{black}{Code smells are indicators of suboptimal design that negatively impact software quality. However, automated code smell detection remains a persistent challenge: heuristics-based tools suffer from high sensitivity to threshold selection and inherent subjectivity, while Machine Learning (ML) and Deep Learning (DL) models yield unsatisfactory performance. Although Large Language Models (LLMs) offer a promising solution to these limitations, their adaptation is impeded by the prohibitive cost of full fine-tuning and the lack of ``LM-ready'' benchmarks, as existing code smell datasets largely rely on software metrics or contain noisy, unverified labels. To address these gaps and enable a rigorous assessment of efficient LLM adaptation for code smell detection, we present a comprehensive study comprising two synergistic contributions. First, we constructed a high-quality, source-code-centric benchmark covering four non-trivial code smells—\textit{Complex Conditional}, \textit{Complex Method}, \textit{Feature Envy}, and \textit{Data Class}—and ensured its reliability through a meticulous two-stage manual review. Second, leveraging this benchmark, we systematically evaluated the effectiveness of four Parameter-Efficient Fine-Tuning (PEFT) methods (i.e., prompt tuning, prefix tuning, LoRA, and $(IA)^3$) across four Small LMs (SLMs) and five LLMs. We benchmarked these PEFT-tuned LMs against a broad suite of baselines, including traditional heuristics-based detectors (\textsc{DesigniteJava} and \textsc{PMD}), DL-based approaches (\textsc{DeepSmells} and \textsc{AE-Dense}), and cutting-edge general-purpose LLMs (GPT-4o-mini and DeepSeek-v3) with In-Context Learning. Our results demonstrate that PEFT methods match or exceed full fine-tuning performance while significantly reducing peak GPU memory usage, and outperform all baselines with MCC improvements ranging from 0.33\% to 13.69\%, particularly for some code smells. Furthermore, we provide actionable insights into the choice of PEFT method for code smell detection, which should be guided by the model used, data availability, and computational resources. Our replication package is available at \cite{replicationPackage} to support reproducibility and future extensions.}
\end{abstract}

\ccsdesc[500]{Software and its engineering~Software maintenance tools}

\keywords{Code Smell Detection, Large Language Model, Parameter-Efficient Fine-Tuning}

\maketitle       

\section{Introduction}
\label{sec:introduction}
\textcolor{black}{Code smells are indicators of underlying design or implementation issues that may degrade software quality \cite{fowler1999refactoring}, particularly in terms of maintainability \cite{palomba2018maintainability}, readability \cite{abbes2011empirical}, and testability \cite{chatzigeorgiou2010investigating}. 
Detecting code smells helps to uncover software design flaws that lead to potential problems in future development and maintenance \cite{fowler1999refactoring, lanza2007object, moha2010decor}. However, manual identification of code smells is often time-consuming and cost-prohibitive \cite{prokic2024prescriptive}, which has prompted extensive research into automated detection techniques.}

\textcolor{black}{Early work predominantly relied on heuristics-based detectors \cite{marinescu2004detection, morales2017on, moha2010decor}. These methods apply manually curated features and thresholds to discriminate between smelly and non-smelly instances \cite{azeem2019machine}. While effective in some settings, such static tools exhibit well-documented shortcomings \cite{fernandes2016review, zhang2011code}: their performance is highly sensitive to threshold selection, they offer limited semantic understanding, and their results often diverge from human judgment due to the inherent subjectivity of code smells \cite{azeem2019machine, fontana2016antipattern, mantyla2006subjective, schumacher2010building}.}

\textcolor{black}{To overcome these limitations, researchers have increasingly incorporated Machine Learning (ML) and Deep Learning (DL) algorithms for code smell detection. These approaches aim to mitigate the performance instability inherent in the manual threshold-setting of heuristics-based detectors \cite{fontana2013code, fontana2016comparing, sharma2021code, liu2021deep}. By reducing reliance on rigid thresholds, ML and DL models can partially learn semantic cues \cite{nucci2018detecting}. Nonetheless, recent studies indicate that these models still struggle to achieve considerable performance \cite{pereira2022code, alazba2023deep}, as they frequently fail to capture the deeper semantic and structural relationships within source code.}

\textcolor{black}{Recently, Large Language Models (LLMs) have emerged as a transformative force in Software Engineering (SE), achieving state-of-the-art performance across a wide range of tasks \cite{lu2023llamareviewer, le2024study}. Despite their impressive general capabilities, directly employing off-the-shelf LLMs for code smell detection without any task-specific fine-tuning yields suboptimal results \cite{mesbah2025leveraging, silva2024detecting, sadik2025benchmarking}. While effective adaptation is essential, two primary challenges currently impede progress:}

\textcolor{black}{\textbf{Challenge 1: Lack of High-Quality, LM-Ready Datasets.} Most existing code smell datasets comprise calculated software metrics rather than the raw source code required by LMs. Moreover, they typically focus on relatively simple smells that do not demand complex semantic reasoning. Furthermore, datasets that do include code often rely on noisy labels generated directly by heuristics-based detectors without human verification, leading to the ``Oracle Problem''. This incompatibility renders existing benchmarks unsuitable for the effective training or evaluation of LMs in code smell detection. Consequently, without high-quality, source-code-centric benchmarks covering non-trivial code smells, a fair investigation into the effectiveness of LMs for code smell detection is infeasible.}

\textcolor{black}{\textbf{Challenge 2: Unclear Effectiveness of LM Adaptation Methods.} Although Parameter-Efficient Fine-Tuning (PEFT) methods have shown promise across various SE tasks by enabling task-specific adaptation with only a small number of trainable parameters \cite{wang2023one, liu2023empirical, he2022towards, liu2024delving, weyssow2023exploring, li2024comprehensive}, their effectiveness for code smell detection remains largely unexplored. It is still unclear whether PEFT approaches can match the performance of full fine-tuning or outperform existing baselines in code smell detection.}

\textcolor{black}{Investigating PEFT methods in the context of code smell detection also fills a broader research gap. Prior studies on PEFT have primarily concentrated on method-level tasks \cite{li2024comprehensive, liu2023empirical, wang2023one, liu2024delving, weyssow2023exploring}, leaving their effectiveness in class-level contexts largely unexplored. Our study addresses this gap by exploring the capabilities of PEFT-adapted LMs in capturing class-level semantics through the detection of code smells, such as \textit{Feature Envy} and \textit{Data Class}, which inherently necessitate class-level code understanding.
Furthermore, utilizing PEFT techniques facilitates just-in-time code smell detection. Traditional heuristics-based detectors are often applied post-development \cite{designitejava, pmd, marinescu2005iplasma}, discovering issues only after software systems are fully implemented and thereby increasing refactoring overhead. In contrast, PEFT-tuned LMs enable real-time identification of code smells, allowing developers to receive immediate feedback upon completing a code block. This proactive approach supports high code quality throughout the development lifecycle, leading to more efficient software maintenance \cite{ardimento2021temporal, ardimento2021transfer}.}

\textcolor{black}{To this end, we systematically evaluate the effectiveness of state-of-the-art PEFT methods on both Small LMs (SLMs, $<$1B parameters) and Large LMs (LLMs, $\ge$1B parameters) for code smell detection. To facilitate this evaluation and resolve the aforementioned data bottleneck, we first curated a high-quality Java code smell dataset covering four types of widely studied and sophisticated code smells at both method and class levels: \textit{Complex Conditional} (\textit{CC}), \textit{Complex Method} (\textit{CM}), \textit{Feature Envy} (\textit{FE}), and \textit{Data Class} (\textit{DC}). Crucially, we subjected this dataset to a rigorous two-stage manual review to guarantee its reliability. With a robust benchmark established, we then addressed the adaptation challenge by fine-tuning four SLMs and five LLMs using various PEFT techniques (i.e., prompt tuning, prefix tuning, LoRA, and (IA)$^3$). We compared these PEFT-tuned LMs against a comprehensive suite of baselines, encompassing heuristics-based detectors, DL-based approaches, and general purpose LLMs with In-Context Learning (ICL).}

\textcolor{black}{The \textbf{main contributions} of this work are summarized as follows:
\begin{itemize}
    \item \textbf{LM-Ready Benchmark for Code Smell Detection:} We construct a high-quality, source-code-centric dataset covering four non-trivial code smells—\textit{Complex Conditional}, \textit{Complex Method}, \textit{Feature Envy}, and \textit{Data Class}. By mining recent Java repositories, applying heuristics-based detectors, and performing a stringent two-stage manual review, we address the limitations of existing code smell datasets (e.g, reliance on code metrics, focusing on simple smells, and unverified labels), ensuring high data quality and a balanced distribution. This establishes a reliable ground truth necessary for training and evaluating LMs for code smell detection, serving as the foundation for our subsequent experiments.
    \item \textbf{Systematic Evaluation of PEFT Methods:} Leveraging our curated dataset, we conduct the first extensive evaluation of PEFT methods for code smell detection on both SLMs and LLMs. We compare multiple PEFT techniques, including prompt tuning, prefix tuning, LoRA, and $(IA)^3$, against full fine-tuning. Our analysis comprehensively assesses their effectiveness, peak GPU memory usage, and performance under different hyper-parameter configurations and low-resource scenarios.
    \item \textbf{Comprehensive Comparison with SOTA Baselines:} To validate the utility of PEFT-tuned LMs for code smell detection, we benchmark these techniques against a wide range of baselines. These baselines span traditional heuristics-based tools, advanced DL-based detectors, and general-purpose LLMs with In-Context Learning under zero-shot, few-shot, and criteria-informed settings. Our results demonstrate that PEFT methods consistently achieve comparable or superior effectiveness while reducing computational cost, highlighting their practical value for the code smell detection.
    \item \textbf{Practical Insights for Developers and Researchers:} Based on our extensive experiments, we distill actionable recommendations for applying PEFT methods to code smell detection. We offer specific guidance on model selection, hyper-parameter tuning strategies, and the trade-offs between efficiency and performance, providing a useful reference for future research and practice in this domain.
\end{itemize}
}


\textcolor{black}{The remainder of this paper is organized as follows: Section \ref{sec:background} introduces the background and motivation. Section \ref{sec:dataset construction} details the dataset construction process, while Section \ref{sec:experiment setup} outlines the research questions and experimental setup. The study results are presented in Section \ref{sec:results}, followed by a discussion in Section \ref{sec:discussion}. Threats to validity are addressed in Section \ref{sec:threats}, with the related work reviewed in Section \ref{sec:related work}. Finally, Section \ref{sec:conclusions} concludes the paper.}

\section{\textcolor{black}{Background and Motivation}}
\label{sec:background}
\subsection{\textcolor{black}{The Case for Applying Language Models on Code Smell Detection}}
\begin{figure}
    \centering
    \includegraphics[width=1\linewidth]{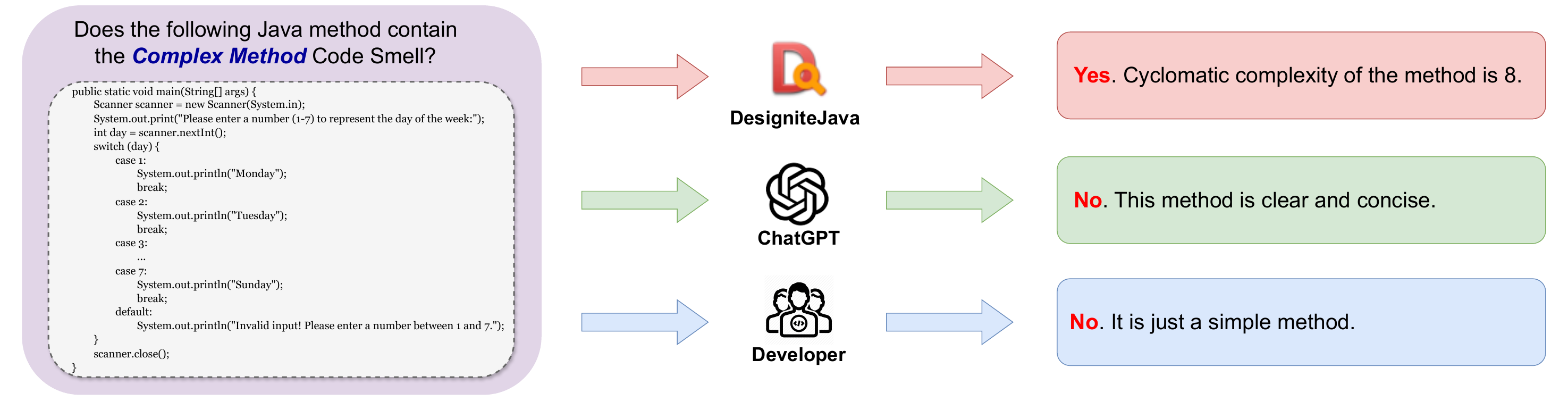}
    \caption{An example of \textit{Complex Method} (\textit{CM}) code smell detection by \textsc{DesigniteJava}, ChatGPT, and human developers}
    \label{fig:example}
\end{figure}

Existing heuristics-based code smell detectors rely on predefined rules and metrics, while ML- or DL-based approaches depend on algorithms to identify the optimal threshold for code smell classification. \textcolor{black}{However, these tools often overlook the semantic and contextual information of code, which is crucial for accurately detecting and understanding code smells in real-world software development \cite{sharma2021code}. For instance, the Java method shown in Figure \ref{fig:example} contains a switch-case statement with seven cases. From the perspective of a Java developer, this method simply maps the numbers 1 through 7 to the days of the week. It is straightforward and easy to understand, as all the case statements follow the same structure. While the method could be refactored to reduce repetition, it has a single responsibility, and its logic is immediately clear to a reader. Therefore, the method does not exhibit a \textit{Complex Method} (\textit{CM}) code smell in practice.
However, when scanned by the heuristics-based code smell detector, \textsc{DesigniteJava} \cite{designitejava}, the method is flagged as \textit{CM} solely because its cyclomatic complexity exceeds a predefined threshold, while its semantic clarity is ignored.}

\begin{figure}[ht]
\centering
\begin{subfigure}[t]{0.49\textwidth}
\begin{lstlisting}[language=Java]
public static String format2Duration(long ms) {
    long days = MILLISECONDS.toDays(ms);
    long hours = MILLISECONDS.toDurationHours(ms);
    long minutes = MILLISECONDS.toDurationMinutes(ms);
    long seconds = MILLISECONDS.toDurationSeconds(ms);

    if (days == 0 && hours == 0 && minutes == 0 && seconds == 0) {
        seconds = 1;
    }

    StringBuilder strBuilder = new StringBuilder();
    if (days > 0) strBuilder.append(days).append("d ");
    if (hours > 0) strBuilder.append(hours).append("h ");
    if (minutes > 0) strBuilder.append(minutes).append("m ");
    if (seconds > 0) strBuilder.append(seconds).append("s");

    return strBuilder.toString();
}
\end{lstlisting}
\caption{False positive: identified as \textit{CC}-positive by \textsc{DesigniteJava} but clean labeled by developers}
\end{subfigure}
\hfill
\begin{subfigure}[t]{0.49\textwidth}
\begin{lstlisting}[language=Java]
public boolean onActionItemClicked(ActionMode mode, MenuItem menuItem) {
    int id = menuItem.getItemId();
    if (id == R.id.delete_entry) {
        confirmDelete(mode);
        return true;
    } else if (id == R.id.select_all) {
        if (mSelected.size() >= mAdapter.getCount()) 
            mode.finish();
        else {
            for (int i = 0; i < mAdapter.getCount(); i++) {
                if (!mListView.isItemChecked(i)) 
                    mListView.setItemChecked(i, true);
            }
        }
        return true;
    } else {
        return false;
    }
}
\end{lstlisting}
\caption{False negative: identified as clean by \textsc{DesigniteJava} but contains developer-labeled CC code smell}
\end{subfigure}
\caption{\textcolor{black}{Examples of \textit{Complex Conditional (CC)} code smell detection by \textsc{DesigniteJava}, highlighting a false positive (a) and a false negative (b)}}
\label{fig:example of CC}
\end{figure}

\textcolor{black}{Figure \ref{fig:example of CC} illustrates additional limitations in rule-based detection for \textit{Complex Conditional} (\textit{CC}) code smells.
Example (a) is identified as \textit{CC}-positive by \textsc{DesigniteJava} but is considered clean by human developers: the conditional \texttt{if (days == 0 \&\& hours == 0 \&\& minutes == 0 \&\& seconds == 0)} exceeds the operand threshold predefined by \textsc{DesigniteJava}, yet the logic is simple and readable, resulting in a false positive.
Example (b) is labeled as clean by \textsc{DesigniteJava} but contains a developer-identified \textit{CC}: nested loops and interacting conditionals make the overall logic difficult to follow, even though no single conditional exceeds structural thresholds, resulting in a false negative.
}

\textcolor{black}{In contrast, Language Models (LMs) such as ChatGPT \cite{achiam2023gpt} can leverage semantic and contextual understanding \cite{wang2021codet5, roziere2024codellama}, enabling detection of both \textit{CM} and \textit{CC} code smells consistent with human judgment. For \textit{CM}, LMs can recognize that a method with high cyclomatic complexity or many lines of code may still be clear and maintainable if its logic is transparent and responsibilities are well-defined. For \textit{CC}, LMs can evaluate not only structural metrics, such as nesting depth or the number of boolean operands, but also semantic aspects such as readability, logical flow, and how the conditions interact, allowing them to correctly identify both false positives and false negatives that rule-based tools might misclassify. By integrating structural and semantic reasoning, LMs offer a more accurate and nuanced approach than traditional heuristics or purely threshold-based ML or DL tools. However, applying LMs on code smell detection effectively requires high-quality data and efficient training strategies, which motivates our work.}


\subsection{\textcolor{black}{The Gap in Code Smell Detection Datasets}}
\label{subsec:the gap in code smell detection datasets}
\textcolor{black}{Despite the growing use of Large Language Models (LLMs) in software engineering research, existing code smell datasets remain a major bottleneck for effective model training. We identify three critical limitations that make current public datasets inadequate for fine-tuning LMs for code smell detection.}

\textcolor{black}{\textbf{1. Predominance of Metric-Based Data}: Traditional code smell detection treated the task as tabular classification, leading to widely used datasets containing only software metrics rather than actual source code \cite{fontana2016comparing, binh2022mlcodesmell, madeyski2023detecting}. While suitable for training classical classifiers like Random Forests or Logistic Regression, these metric-based datasets are incompatible with LMs, which require tokenized source code to capture semantic and syntactic patterns.}

\textcolor{black}{\textbf{2. Limited Scope of Code Smell Types:} Existing datasets often lack diversity in the types of smells covered \cite{fontana2016comparing, madeyski2020mlcq, fontana2017code, liu2021deep, moha2009decor}. There is a disproportionate focus on easily detectable, metric-based smells (e.g., \textit{Long Method}). Conversely, there is a scarcity of high-quality samples for structurally intricate smells, such as \textit{Complex Conditional} and \textit{Complex Method}, which necessitate semantic reasoning beyond simple metric thresholds.}

\textcolor{black}{\textbf{3. The ``Oracle Problem'' and Label Reliability:} A significant number of large-scale datasets rely on heuristics-based detectors (e.g., \textsc{PMD}, \textsc{DECOR}, and \textsc{iPlasma}) as the ``oracle'' to automatically generate labels, typically bypassing manual verification \cite{moha2009decor, lenarduzzi2019technical, sae2016context, yang2015classification, khomh2012exploratory}. However, the inherent inaccuracies of these tools introduce noise into the ground truth \cite{palomba2018diffuseness, de2017towards}. 
Training an LM on such datasets risks merely teaching the model to emulate the flaws and biases of heuristics-based detectors rather than identifying genuine design issues.}

\textcolor{black}{In summary, current code smell datasets are not ``LM-ready''. To effectively fine-tune an LM for code smell detection, there is an urgent need for a dataset that provides raw, compilable source code, covers complex semantic smells, and relies on high-quality, human-validated labels. This motivates the construction of our proposed dataset.}

\subsection{\textcolor{black}{Full Fine-Tuning and Parameter-Efficient Fine-Tuning (PEFT)}}
\subsubsection{\textcolor{black}{Full Fine-Tuning}}
Language Models (LMs) which are pre-trained on massive general-purpose corpus via self-supervised method, can be applied to various downstream NLP tasks by adapting their rich possession of learned knowledge through full fine-tuning \cite{qiu2020pre}. Full fine-tuning involves updating all parameters of an LM, including multi-head attention (MHA), feed-forward neural network (FFN), layer normalization, and so on. However, as LMs scale up, full fine-tuning a large-sized LM, i.e., LLM, can be excessively costly and ultimately make it practically unfeasible \cite{hu2024survey, ding2023parameter}.

\subsubsection{\textcolor{black}{Parameter-Efficient Fine-Tuning (PEFT) Methods}}
\label{subsec:explainations of peft}
Instead of full fine-tuning, pre-trained LMs could be tailored to downstream tasks by Parameter-Efficient Fine-Tuning (PEFT) methods. PEFT methods involve optimizing only a subset of model parameters or incorporating external modules for new tasks while keeping the majority of parameters frozen. Compared to full fine-tuning, PEFT methods can reduce the substantial computational and data demands of LLMs and achieve comparable or even better results \cite{han2024parameter}. 

\textcolor{black}{In this study, we employed the Hugging Face PEFT library\footnote{\url{https://github.com/huggingface/peft}} to implement and evaluate all PEFT methods. This library, developed by the Hugging Face team, offers a standardized and efficient framework that supports several widely adopted PEFT techniques and has been extensively used in recent research \cite{li2024comprehensive, weyssow2023exploring, wang2025beyond, storhaug2024parameter}. In addition, it is fully compatible with the Transformers library, allowing seamless integration with open-source LMs available on the Hugging Face platform. This unified setup ensures reproducibility, simplifies implementation, and enables fair comparisons across different PEFT strategies. Therefore, in this work, we focus exclusively on the PEFT methods provided by the PEFT library. The right side of Figure \ref{fig:architecture and illustration} illustrates the four state-of-the-art PEFT techniques adopted in our study: prompt tuning, prefix tuning, LoRA, and $(IA)^{3}$.} 

\textcolor{black}{These methods were selected based on several considerations. First, they are all natively supported by the PEFT library, allowing stable fine-tuning of LMs while avoiding out-of-memory errors. Second, according to the official documentation\footnote{\url{https://huggingface.co/docs/peft/index}}, PEFT methods are categorized into three main types: Prompt-based methods, LoRA methods, and $(IA)^3$. The four selected methods span these categories, making them representative of the major approaches currently available. Finally, these methods are not only recent but also widely adopted in contemporary research \cite{weyssow2023exploring, li2024comprehensive, mannisto2025comparative, zhou2024empirical, taylor2022galactica, esmaeili2024empirical, wistuba2024choice}. Their demonstrated effectiveness across various tasks and their popularity in both the NLP and SE communities make them ideal candidates for this study. By selecting these four methods, we ensure that our experiments cover a range of the most promising and commonly adopted techniques in PEFT.}

\textbf{Prompt tuning} \cite{lester2021power} directly appends learnable vectors, known as soft prompts, to the initial input embedding layer. When fine-tuning PLMs with prompt tuning, only the soft prompts are updated, while the remaining model weights and architecture are kept fixed (not updated).

\textbf{Prefix tuning} \cite{li2021prefix} introduces adjustable prefix (i.e., a sequence of task-specific vectors) across all Transformer layers. Specifically, these trainable prefix vectors are prepended to the key $K$ and value $V$ of all the MHA mechanisms. Prefix tuning adopts a reparameterization strategy, utilizing a multilayer perceptron (MLP) layer to generate the added vectors instead of optimizing them in a straightforward manner \cite{han2024parameter}.

\textbf{LoRA} (Low-Rank Adaptation) \cite{hu2022lora} injects trainable low-rank decomposition matrices into the query $W_q$, key $W_k$, and value $W_v$ projection matrices of the MHA. For a pre-trained weight matrix $W_0 \in \{W_q, W_k, W_v\}$, an additive operation $W_0 + \Delta W = W_0 + BA$ is performed, where the rank of matrices $B$ and $A$ is considerably smaller than that of $W_0$, consequently effectively reducing the number of trainable parameters while preserving the essential characteristics of the original matrices.

$\mathbf{(IA)^{3}}$ \cite{liu2022few} (Infused Adapter by Inhibiting and Amplifying Inner Activation) incorporates three learnable rescaling vectors, $l_k$, $l_v$, and $l_{ff}$, to rescale the key $K$ and value $V$ in MHA and the inner activations in FFN. The operations conducted within the MHA can be outlined as follows:
\begin{equation}
    \text{MHA(x)} = Softmax\Biggl(\frac{Q(l_k \odot K^T)}{\sqrt{d_k}}\Biggr)(l_v \odot V)
\end{equation}
where $\odot$ denotes the element-wise multiplication and $d_k$ refers to the dimension of key $K$.

The modifications in the FFN can be defined as:
\begin{equation}
    \text{FFN(x)} = W_{up}(l_{ff} \odot \sigma(W_{down}x))
\end{equation}
where $\sigma$ is the activation function within FFN, and $W_{up}$ and $W_{down}$ indicate the weight matrices of the FFN.


\section{\textcolor{black}{Dataset Construction}}
\label{sec:dataset construction}
\textcolor{black}{In this section, we elaborate on the process of constructing high-quality and balanced datasets for four types of code smells, rendering them LM-ready and tailored specifically for fine-tuning both small and large LMs.}

\textcolor{black}{As discussed in Section \ref{subsec:the gap in code smell detection datasets}, existing code smell datasets suffer from several critical limitations, including the absence of raw source code, insufficient coverage of complex smells, and label noise. To address these issues and ensure reproducibility, we do not rely on any existing datasets. Instead, we create our benchmark from scratch by mining recent Java repositories from GitHub, providing a solid foundation for the subsequent fine-tuning experiments on code smell detection.}

\begin{figure}
    \centering
    \includegraphics[width=1\linewidth]{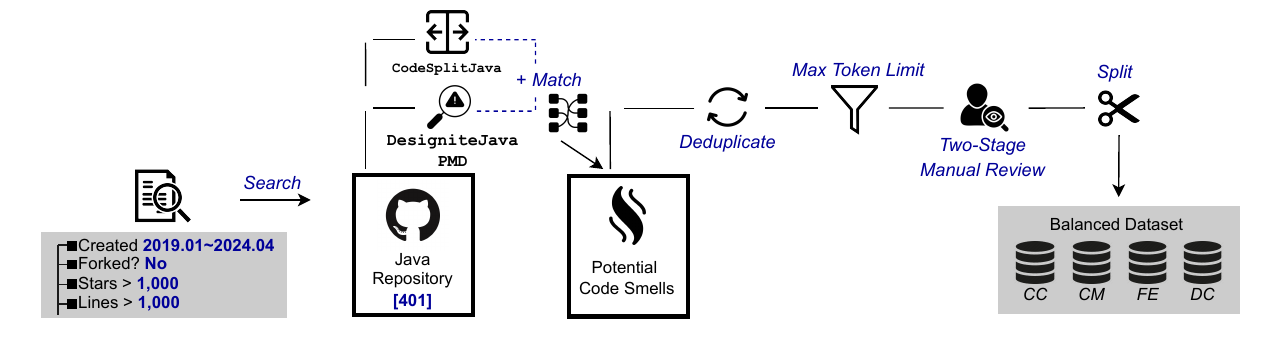}
    \caption{Overview of dataset construction}
    \label{fig:Overview of dataset construction}
\end{figure}

\textcolor{black}{Our construction pipeline is guided by three key objectives: (1) Source-Code Centricity: ensuring all samples contain compilable raw code tokens; (2) Diversity: covering four non-trivial code smells (i.e., \textit{Complex Conditional}, \textit{Complex Method}, \textit{Feature Envy}, \textit{Data Class}); and (3) Reliability: incorporating a rigorous two-stage manual review to alleviate the ``Oracle Problem''. Figure \ref{fig:Overview of dataset construction} provides an overview of the dataset construction process.}

\subsection{Selected Code Smells}
\label{subsubsec:selected code smells}
In this study, we intend to use LMs with PEFT methods to detect four types of code smells, comprising three at the method level (i.e., \textit{Complex Conditional}, \textit{Complex Method}, and \textit{Feature Envy}) and one at the class level (i.e., \textit{Data Class}). The definitions of these four code smell types are provided below:
\begin{itemize}
    \item \textbf{\textit{Complex Conditional} (\textit{CC}):} This code smell occurs when a conditional statement is excessively intricate \cite{sharma2017house}.
    \item \textbf{\textit{Complex Method} (\textit{CM})}: This code smell occurs when a method exhibits high cyclomatic complexity \cite{sharma2017house}.
    \item \textbf{\textit{Feature Envy} (\textit{FE}):} This code smell occurs when a method is more interested in a class other than the one it actually is in \cite{fowler1999refactoring, fontana2016comparing}.
    \item \textbf{\textit{Data Class} (\textit{DC})}: This code smell occurs when a class contains only fields and basic methods for accessing them \cite{fowler1999refactoring, fontana2016comparing}.
\end{itemize}

We selected \textit{CC}, \textit{CM}, \textit{FE}, and \textit{DC} because they are among the most extensively studied code smells in prior research on ML- and DL-based code smell detection (e.g., \cite{sharma2021code, ho2023fusion, zhang2024code, alazba2023deep, fontana2016comparing, barbez2020machine, kim2017finding, liu2023deep, liu2018deep, skipina2024automatic, yang2024feature, fontana2013code, madeyski2020mlcq}). Moreover, these code smells are frequently observed in real-world software projects and are known to significantly impact code quality, maintainability, and the likelihood of introducing defects \cite{fowler1999refactoring}. For example, \textit{CC} and \textit{CM} can lead to convoluted control flows that hinder code readability and increase cognitive load on developers, while \textit{FE} and \textit{DC} often indicate bad modularization and poor class design, which complicate maintenance and evolution \cite{fowler1999refactoring, sjoberg2012questioning, khomh2009exploratory}.

Another reason for choosing these specific smell types is the availability of detection tools, such as \textsc{DesigniteJava} \cite{designitejava} and \textsc{PMD} \cite{pmd}, which facilitates the collection of candidate code samples potentially exhibiting these smells. We excluded overly simple smells (e.g., \textit{Long Method}) that can be easily identified using basic heuristics. In contrast, detecting the four selected code smells requires LLMs to capture subtle features, consider aspects such as cohesion and coupling, and analyze the overall structure of a code sample \cite{sharma2021code}, which expands the ambition and enhances the applicability of this study.

At the same time, identifying these selected code smells allows us to assess the effectiveness of LMs fine-tuned with PEFT methods in both method-level and class-level code understanding. \textit{CC}, \textit{CM}, and \textit{FE} are generally considered method-level code smells, while \textit{DC} represents a class-level code smell. However, detecting \textit{FE} requires comprehension of not only the focal method but also its surrounding class context and inter-class dependencies \cite{fowler1999refactoring, tsantalis2009indentification, liu2023deep, alkharabsheh2021analysing}. Therefore, it also demands class-level code understanding. The selection of these four code smells allows our study to address a key gap in existing research related to PEFT methods \cite{li2024comprehensive, liu2023empirical, wang2023one, liu2024delving, weyssow2023exploring}, which has primarily concentrated on the effectiveness of PEFT techniques on method-level code-related tasks.

\subsection{Java Repository Selection}
\label{subsubsec: Java Repository Selection}
To construct a code smell dataset that accurately represents modern coding practices, we selected Java repositories created in the past six years from GitHub. We used the GitHub REST API\footnote{\url{https://docs.github.com/en/rest?apiVersion=2022-11-28}} to gather available Java projects that meet the following criteria:
\begin{itemize}
    \item The repositories should be created between January 1, 2019 and April 30, 2024.
    \item The repositories should exclude forked ones to avoid duplicates.
    \item The repositories should have a minimum of 1,000 stars and 1,000 lines of code.
\end{itemize}

These criteria ensured that the selected Java repositories were recent, popular, active, and non-trivial. By employing these stringent filtering criteria, we obtained a list of 401 GitHub repositories coded in Java.

\subsection{Potential Code Smell Detection}
\label{subsec: Code Smell Detection}
\textsc{DesigniteJava} \cite{designitejava} is an open-source tool capable of detecting 18 design smells and 10 implementation smells, whereas \textsc{PMD} \cite{pmd} is a rule-based static analysis tool that identifies common programming issues, including code smells. Both of them have been extensively used in prior studies \cite{sharma2021qscored, shah2023mining, barbosa2020revealing, fernandes2016review, aniche2018code, wu2024ismell}. In this paper, we leveraged \textsc{DesigniteJava} to detect potential instances of \textit{CC}, \textit{CM}, and \textit{FE}, and used \textsc{PMD} to detect potential instances of \textit{DC}. The output of this step is a list of detected code smells, each annotated with the corresponding project, package, class, and, where applicable, method name. It is worth noting that the code smell detection results from \textsc{DesigniteJava} and \textsc{PMD} may not be entirely accurate, as the detection results can contain both false positives and false negatives. False positives occur when the tool incorrectly flags code as exhibiting certain type of code smell, while false negatives happen when actual code smells are missed. These inaccuracies stem from the rule-based and heuristic nature of \textsc{DesigniteJava} and \textsc{PMD}, which relies on predefined criteria that may not capture all nuances of the code.

\subsection{Data Preparation}
\label{subsubsec:Data preparation}
\textsc{CodeSplitJava} \cite{codesplitjava} is a utility program designed to extract methods or classes from Java source code and organize them into separate files. We employed \textsc{CodeSplitJava} to parse the 401 Java repositories collected in Section~\ref{subsubsec: Java Repository Selection}, generating individual code files for each method or class. The file paths reflect the hierarchical structure of the original code (i.e., namespaces and packages are converted into nested folders). By aligning the list of detected code smells obtained in Section \ref{subsec: Code Smell Detection} with the paths of the code files generated by \textsc{CodeSplitJava}, we were able to collect code snippets that were labeled by heuristics-based detectors as either containing code smells or being free of them.

\begin{table}[htbp]
\centering
\footnotesize
\caption{Number of potential positive code samples after each step of data preparation (\textbf{\#Initial Samples} shows the initially collected potential positive samples, \textbf{\#After Deduplication} reflects the number of samples after duplicate removal, and \textbf{\#After Token Limit} indicates the count of samples that pass the token length filter, i.e., no more than 1,024 tokens)}
\label{tab:Number of code snippets after each step of data preparation}
\begin{tabular}{m{2.6cm}|m{3cm}<{\centering\arraybackslash}|m{3cm}<{\centering\arraybackslash}|m{3cm}<{\centering\arraybackslash}}
\toprule
\textbf{Code Smell} & \textbf{\#Initial Samples} & \textbf{\#After Deduplication} & \textbf{\#After Token Limit} \\ \cmidrule{1-4}
Complex Conditional & 23,984                     & 7,768                          & 6,901 \\ \cmidrule{1-4}
Complex Method      & 20,171                     & 14,687                         & 9,819 \\ \cmidrule{1-4}
Feature Envy        & 2,152                      & 1,113                          & 337   \\ \cmidrule{1-4}
Data Class          & 9,672                      & 8,501                          & 7,121 \\ 
\bottomrule
\end{tabular}
\end{table}

We removed any duplicate samples within each type of code smell to ensure uniqueness. Additionally, we limited the maximum token length of the samples to 1,024, filtering out any that exceeded this limit. This decision was made to accommodate the input constraints of the models used in our experiments (see Section \ref{subsec:implementation details}). Specifically, the input window size for the LMs was set to 1,024 tokens to align with GPU memory limitations. Samples exceeding 1,024 tokens undergo truncation, which may result in the loss of critical contextual information. For instance, key sections of code relevant to the detection of code smells could be omitted, potentially impacting the model's performance. Therefore, we set a hard limit of 1,024 tokens to avoid such issues, ensuring that all samples in our dataset can be fully utilized by the models without sacrificing the integrity of their content.
Table \ref{tab:Number of code snippets after each step of data preparation} provides the number of potential positive samples after each step for the four types of code smells. Note that although \textit{FE} is a method-level code smell, we retained the entire class that contains the method as the sample. This decision aligns with the definition of \textit{FE} (see Section \ref{subsubsec:selected code smells}), which emphasizes the method's interaction with its enclosing class. Providing the full class context ensures that the code sample contains sufficient structural and semantic information for \textit{FE} detection. 

\subsection{Manual Review}
\label{subsec: manual review}
\begin{figure}
    \hspace{1.25cm}
    \includegraphics[width=0.8\linewidth]{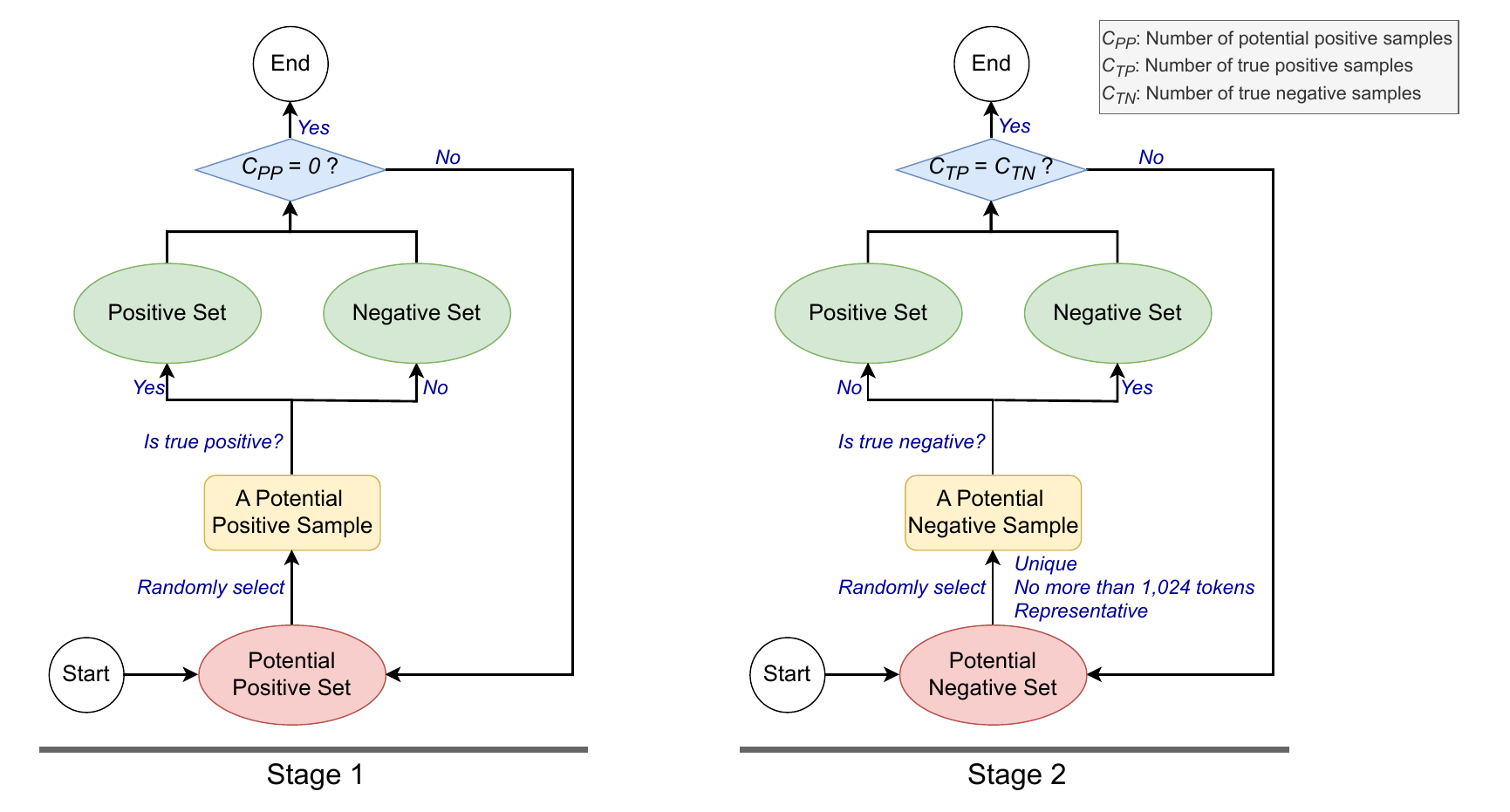}
    \caption{The process of two-stage manual review in dataset construction}
    \label{fig:manual review}
\end{figure}

After the above steps, we obtained two sets of samples for each code smell: a potential positive set and a potential negative set, with the latter being significantly larger in size. To construct a high-quality and balanced dataset, the first and the fourth authors carried out a two-stage manual review process, as illustrated in Figure \ref{fig:manual review}.

In the first stage, we reviewed all potential positive samples manually. Each sample was carefully examined to identify and eliminate false positives. If a sample in the potential positive set had been incorrectly labeled as positive by \textsc{DesigniteJava} or \textsc{PMD}, it was reclassified as negative and moved to the negative set; otherwise, it was transferred into the positive set. This stage yielded a verified positive set and a smaller verified negative set. 
In the second stage, in order to enlarge the negative set to achieve dataset balance, we randomly selected samples from the potential negative set for manual validation. Similar to the first stage, samples confirmed as true negatives were added to the negative set, while those identified as false negatives were reclassified into the positive set. This process was repeated iteratively until an equal number of manually reviewed positive and negative samples was achieved. During this stage, all randomly selected potential negative samples were required to be unique, adhere to the predefined maximum token limit, and be representative. To ensure the representativeness of negative samples across the four types of code smells, we established the following criteria: for \textit{CC}, negative samples must contain at least one conditional statement; for \textit{CM}, negative samples must contain at least 15 lines of code; and for \textit{FE} and \textit{DC}, negative samples must contain at least 30 lines of code, as we extract the corresponding class for these two smells. These constraints help filter out trivial or uninformative code snippets and ensure that the negative samples are sufficiently complex and realistic, thereby contributing to a more robust and reliable dataset.
Through this two-stage review, we ensured that the final dataset for each code smell type was both balanced and of high quality.

\begin{table}[htbp]
\centering
\footnotesize
\textcolor{black}{
\caption{\textcolor{black}{Number of positive and negative samples at each stage of manual review}}
\label{tab:Number of positive and negative samples at each stage of manual review}
\begin{tabular}{m{2.4cm}|m{1.2cm}<{\centering\arraybackslash}|m{1.2cm}<{\centering\arraybackslash}|m{1.2cm}<{\centering\arraybackslash}|m{1.2cm}<{\centering\arraybackslash}|m{1.2cm}<{\centering\arraybackslash}|m{1.2cm}<{\centering\arraybackslash}}
\toprule
\multirow{2}{*}{\textbf{Code Smell}} & \multicolumn{2}{c|}{\textbf{Stage 1}} & \multicolumn{2}{c|}{\textbf{Stage 2}} & \multicolumn{2}{c}{\textbf{Total}} \\ \cmidrule{2-7}
& \textbf{\#Positive} & \textbf{\#Negative} & \textbf{\#Positive} & \textbf{\#Negative} & \textbf{\#Positive} & \textbf{\#Negative}  \\ \cmidrule{1-7}
Complex Conditional   & 3,715               & 3,186               & 889               & 1,418               & 4,604               & 4,604                       \\ \cmidrule{1-7}
Complex Method        & 5,699               & 4,120               & 49                & 1,628               & 5,748               & 5,748                     \\ \cmidrule{1-7}
Feature Envy          & 184                 & 153                 & 123               & 154                 & 307                 & 307                         \\ \cmidrule{1-7}
Data Class            & 5,904               & 1,217               & 1,018             & 5,705               & 6,922               & 6,922                       \\ 
\bottomrule
\end{tabular}
}
\end{table}


In order to ensure the consistency of data annotation, a pilot review was performed by the first and the fourth authors before conducting the formal manual review process. For each code smell type (i.e., \textit{CC}, \textit{CM}, \textit{FE}, and \textit{DC}), 30 potential positive samples and 30 potential negative samples labeled by heuristics-based tools were randomly selected. The two authors independently examined these samples and then compared their labeling results. Cohen’s Kappa coefficient \cite{jacob1960coefficient} was used to assess inter-rater agreement between the two authors. The Kappa coefficient for \textit{CC} was 0.83, for \textit{CM} was 0.80, for \textit{FE} was 0.77, and for \textit{DC} was 0.87. All values exceeded the threshold of 0.7, indicating a strong level of agreement between the annotators. Following the pilot review, the two authors held a discussion to reconcile discrepancies and formalize the labeling criteria below:
\begin{itemize}
    \item \textbf{\textit{CC}:} \textcolor{black}{Labeled as positive if the method contains at least one conditional statement that satisfies (i) a nesting depth of three or more or (ii) compound boolean expressions involving three or more operands, and simultaneously (iii) exhibits semantically complex logic that makes the condition difficult to evaluate and understand.}
    \item \textbf{\textit{CM}:} Labeled as positive if the method simultaneously satisfies both (i) three or more execution paths with complex control flow, and (ii) an implementation that is difficult to comprehend or follow.
    \item \textbf{\textit{FE}:} Labeled as positive if the method inside the input class satisfies both (i) over 50\% of its data access operations, such as method calls and field accesses, target members of external classes rather than its own class, and (ii) relocating the method to one of these external classes would improve encapsulation and cohesion.
    \item \textbf{\textit{DC}:} Labeled as positive if the class contains (i) primarily fields accompanied by trivial getters/setters (at least 80\% of its fields), and (ii) lacks methods that implement significant logic or domain behavior, indicating its role as a passive data holder.
\end{itemize}

These criteria were grounded in established definitions from prior studies \cite{sharma2017house, fowler1999refactoring, fontana2016comparing, pietrzak2006leveraging} and refined based on insights from our pilot study. During the formal manual review, the two authors collaboratively labeled all samples (i.e., each sample was assigned to one of the two authors), which took approximately 300 person-hours in total. Whenever questions or ambiguities arose, the second author, who is an expert in code smell detection, was consulted. In such cases, all three authors discussed the sample until consensus was reached, ensuring annotation reliability and consistency. \textcolor{black}{Table \ref{tab:Number of positive and negative samples at each stage of manual review} summarizes the numbers of positive and negative samples collected at each stage of the manual review process.}

\subsection{Data Split}
\label{subsec:data split}
In this step, we split the code samples of \textit{CC}, \textit{CM}, \textit{FE}, and \textit{DC} into training, validation, and test sets following an 8:1:1 ratio, as employed in previous studies \cite{yang2024federated, liu2023benchmarking}. For each type of code smell, we ensured that the number of positive and negative samples remains equal across the training, validation, and test sets, thereby maintaining class balance throughout the entire dataset.
Table \ref{tab:Number of positive and negative samples after data split} outlines the number of code samples for each code smell type after the data split. We then used these clean and high-quality datasets as our final datasets for a series of experiments.

\begin{table}[htbp]
\centering
\footnotesize
\caption{Number of positive and negative samples after data split}
\label{tab:Number of positive and negative samples after data split}
\begin{tabular}{m{2.4cm}m{1.2cm}<{\centering\arraybackslash}|m{1.2cm}<{\centering\arraybackslash}|m{1.2cm}<{\centering\arraybackslash}|m{1.2cm}<{\centering\arraybackslash}|m{1.2cm}<{\centering\arraybackslash}}
\toprule
\multicolumn{2}{c|}{\textbf{Code Smell}} & \textbf{Train} & \textbf{Valid} & \textbf{Test} & \textbf{Total} \\ \cmidrule{1-6}
\multirow{2}{*}{Complex Conditional}     & \#Positive & 3,684 & 460 & 460 & 4,604 \\ \cmidrule{2-6}
                                         & \#Negative & 3,684 & 460 & 460 & 4,604\\ \cmidrule{1-6}
\multirow{2}{*}{Complex Method}          & \#Positive & 4,600 & 574 & 574 & 5,748\\ \cmidrule{2-6}
                                         & \#Negative & 4,600 & 574 & 574 & 5,748\\ \cmidrule{1-6}
\multirow{2}{*}{Feature Envy}            & \#Positive & 245 & 31 & 31 & 307 \\ \cmidrule{2-6}
                                         & \#Negative & 245 & 31 & 31 & 307 \\ \cmidrule{1-6}
\multirow{2}{*}{Data Class}              & \#Positive & 5,538 & 692 & 692 & 6,922 \\ \cmidrule{2-6}
                                         & \#Negative & 5,538 & 692 & 692 & 6,922 \\
\bottomrule 
\end{tabular}
\end{table}

\section{\textcolor{black}{Experimental Setup}}
\label{sec:experiment setup}

\subsection{\textcolor{black}{Task Formulation}}
\textcolor{black}{We formulate code smell detection as a binary classification task. Given a code snippet $\mathcal{C} = {c_1, c_2, \dots, c_n}$ consisting of a sequence of tokens, the goal is to learn a mapping function $f: \mathcal{C} \to \mathcal{Y}$, where $\mathcal{Y} \in {0, 1}$. Here, $1$ indicates the presence of a specific code smell, and $0$ denotes its absence.}

\begin{figure}
    \centering
    \includegraphics[width=1\linewidth]{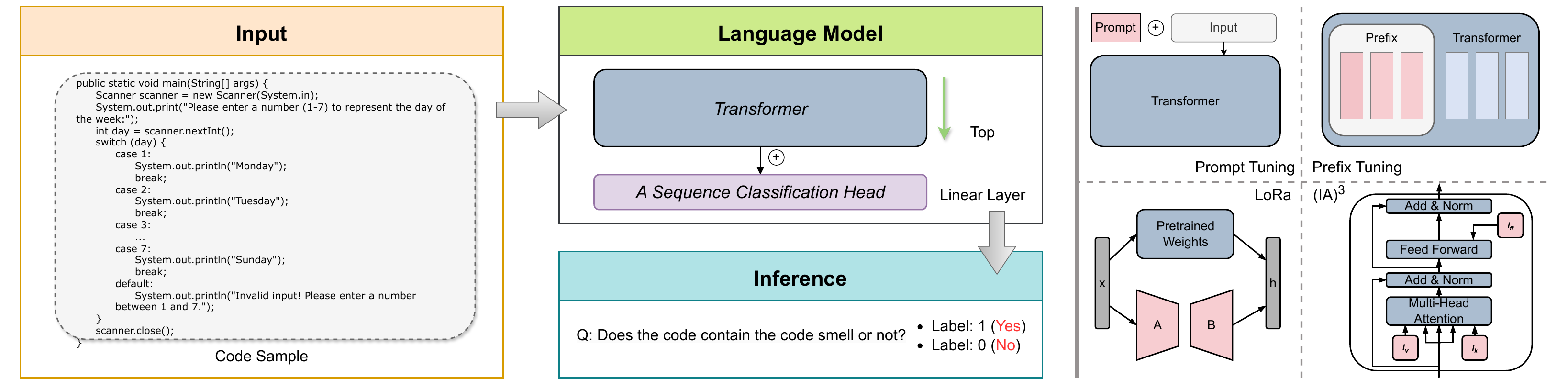}
    \caption{Left: architecture of our proposed method for code smell detection. Right: a detailed illustration of the PEFT module structure for the four PEFT methods.}
    \label{fig:architecture and illustration}
\end{figure}

\textcolor{black}{As illustrated in Figure \ref{fig:architecture and illustration}, we employ a Language Model (LM) as the backbone for this function $f$. To adapt the LM for this specific classification task, we append a sequence classification head on top of the model and optimize it using four Parameter-Efficient Fine-Tuning (PEFT) techniques.}

\subsection{Research Questions}
\label{subsec:research questions}
This study aims to evaluate the effectiveness of state-of-the-art PEFT methods (prompt tuning, prefix tuning, LoRA, and $(IA)^{3}$) in SLMs and LLMs for code smell detection. To achieve this goal, we conducted a set of experiments to answer the following Research Questions (RQs):

\textbf{RQ1: What is the effectiveness of different PEFT methods and full fine-tuning on various SLMs and LLMs?}
Previous studies have shown remarkable results using PEFT methods to fine-tune SLMs for software engineering tasks like code clone detection and defect detection \cite{liu2023empirical, liu2024delving, li2024comprehensive, wang2023one, weyssow2023exploring}. However, these studies have not paid attention to the effectiveness of PEFT methods for code smell detection. With this RQ, we aim to compare and validate the effectiveness of PEFT methods across multiple LMs of varying parameter sizes (see Section \ref{sec:base model selection}) specifically for code smell detection.

\textbf{RQ2: What is the effectiveness of PEFT methods with different hyper-parameters?}
When fine-tuning LMs for downstream tasks, the configuration of hyper-parameters plays a crucial role in determining the final results by affecting the number of trainable parameters. In prompt tuning and prefix tuning, the number of virtual tokens serves as a critical hyper-parameter \cite{lester2021power, li2021prefix}. For LoRA, the key hyper-parameter is the rank $r$ \cite{hu2022lora}, while in $(IA)^3$, it is the selection of activated modules \cite{liu2022few}. Due to our limited computational resources and the popularity of LoRA compared to other PEFT methods \cite{gupta2024beyond, zhang2024star}, we decided to employ LoRA to fine-tune a SLM and an LLM with different $r$ values in a range of \{8, 32, 64, 640, 1,280, 2,560\} to explore the effect of hyper-parameter settings on PEFT methods for code smell detection.

\textbf{RQ3: What is the effectiveness of PEFT methods in low-resource scenarios?}
Low-resource scenarios, characterized by constraints on the amount of training data, pose significant challenges to fine-tuning LMs for specific tasks. Due to the scarcity of real-world datasets \cite{zakeri2023systematic}, these scenarios are common in tasks such as code smell detection. Therefore, it is crucial to investigate how the number of training samples influences the effectiveness of PEFT methods for code smell detection.

\textbf{RQ4: What is the effectiveness of PEFT methods compared to state-of-the-art code smell detection approaches?}
Existing code smell detectors include traditional heuristics-based and DL-based methods. Additionally, the use of general-purpose LLMs with In-Context Learning (ICL) has become increasingly popular in SE tasks \cite{zhou2024large, shin2023prompt, li2025large, xu2024unilog}, and such models can also be leveraged for code smell detection. With this RQ, we aim to assess how LMs utilizing PEFT methods compare against heuristics-based code smell detectors, established DL techniques, and general-purpose LLMs under zero- and few-shot settings for the task of code smell detection.

\subsection{Base Model Selection}
\label{sec:base model selection}
In this work, we aim to explore the effectiveness of PEFT methods for the task of code smell detection. To ensure a thorough analysis, we included both small and large LMs as our base models. Consistent with the definition adopted by Weyssow \textit{et al.} \cite{weyssow2023exploring}, we classified models with 1B or more parameters as LLMs, and those with fewer parameters as SLMs.

\begin{table}[htbp]
\centering
\footnotesize
\caption{\textcolor{black}{Selected LMs and their corresponding parameter sizes, context window limits, and architecture types}}
\label{tab:Selected LLMs and their corresponding parameter sizes, context windows, and architecture types}
\begin{tabular}{m{1.4cm}<{\centering\arraybackslash}m{2.8cm}<{\arraybackslash}|m{2cm}<{\centering\arraybackslash}|m{2.55cm}<{\centering\arraybackslash}|m{2cm}<{\arraybackslash}}
\toprule
\multicolumn{2}{c|}{\textbf{Model}}                  & \textbf{\#Params}  & \textcolor{black}{\textbf{Context Window}} & \textbf{Architecture}\\
\cmidrule{1-2} \cmidrule{3-3} \cmidrule{4-5}
\multirow{4}{*}{\textit{SLMs}} & CodeBERT            & 125M               & \textcolor{black}{512}                     & \multirow{3}{*}{Encoder-only} \\ 
                               & GraphCodeBERT       & 125M               & \textcolor{black}{512} \\ 
                               & UnixCoder           & 125M               & \textcolor{black}{1,024} \\ \cmidrule{2-5}
                               & CodeT5              & 220M               & \textcolor{black}{512}                     & Encoder-decoder \\
\midrule
\midrule
\multirow{5}{*}{\textit{LLMs}} & StarCoderBase-1B    & 1.1B               & \textcolor{black}{1,024}                   & \multirow{5}{*}{Decoder-only} \\
                               & StarCoderBase-3B    & 3B                 & \textcolor{black}{1,024} \\
                               & DeepSeek-Coder-1.3B & 1.3B               & \textcolor{black}{1,024} \\
                               & DeepSeek-Coder-6.7B & 6.7B               & \textcolor{black}{1,024} \\
                               & CodeLlama-7B        & 6.7B               & \textcolor{black}{1,024} \\
\bottomrule
\end{tabular}
\end{table}

First, the selected LMs should be open-source as PEFT methods require access to their parameters. 
Second, the models should support classification tasks as we frame code smell detection as a binary classification problem.
Third, the base models used in this study must be able to be fine-tuned and tested on a single 40GB GPU (our maximum computational resource) without encountering memory overflow.
Fourth, the selected models should either be popular SLMs or recent advanced LLMs to ensure they represent state-of-the-art capabilities.

Based on the above criteria, we employed 4 SLMs (i.e., CodeBERT \cite{feng2020codebert}, GraphCodeBERT \cite{guo2021graphcodebert}, UnixCoder \cite{guo2022unixcoder}, and CodeT5 \cite{wang2021codet5}) and 5 LLMs (StarCoderBase-1B \cite{li2023starcoder}, StarCoderBase-3B \cite{li2023starcoder}, DeepSeek-Coder-1.3B \cite{guo2024deepseekcoder}, DeepSeek-Coder-6.7B \cite{guo2024deepseekcoder}, and CodeLlama-7B \cite{roziere2024codellama}) for our experiments. 
To ensure a comprehensive analysis, we deliberately included models representing all three major Transformer architectures: encoder-only, encoder-decoder, and decoder-only. Although decoder-only models were originally designed for generative tasks \cite{vaswani2017attention, radford2019language}, prior studies have shown their competitive performance in sequence classification tasks as well \cite{wolf2020transformers, kariyappa2024progressive}. Therefore, we also included decoder-only models as base models in our study. Additionally, all the models we selected have demonstrated encouraging results across various code understanding tasks like code clone detection, defect detection, etc. \cite{liu2024vul, karmakar2021what, lu2021codexglue, li2022automating, tang2023csgvd, zeng2022an, chen2023diversevul, wang2023one, fu2022vulrepair, nguyen2024empirical, su2024distilled}. \textcolor{black}{Table \ref{tab:Selected LLMs and their corresponding parameter sizes, context windows, and architecture types} presents the details of the 9 base models, including their parameter sizes, context window limits, and architectural types. The input sequences of CodeBERT, GraphCodeBERT, and CodeT5 were truncated at 512 tokens, which corresponds to their maximum supported input length. For all other models, inputs were uniformly truncated at 1,024 tokens.}

\subsection{\textcolor{black}{Baselines}}
\label{subsec:baselines}
\textcolor{black}{In RQ4, to comprehensively evaluate the effectiveness of PEFT-tuned LMs for code smell detection, we compare them against three distinct categories of baselines: traditional heuristics-based detectors, existing Deep Learning (DL)-based approaches, and state-of-the-art general-purpose LLMs using In-Context Learning (ICL).}

\subsubsection{\textcolor{black}{Heuristics-based detectors}}
\textcolor{black}{We select two prominent heuristics-based detectors that rely on predefined metrics and rule-based algorithms to identify code smells:
\begin{itemize}
    \item \textbf{\textsc{DesigniteJava}} \cite{designitejava}\textbf{:} An industrial-strength tool that detects architecture, design, and implementation smells by analyzing source code structure and calculating metrics (e.g., cyclomatic complexity and class coupling). In this study, we deploy \textsc{DesigniteJava} to detect \textit{CC}, \textit{CM}, and \textit{FE}. 
    \item \textbf{\textsc{PMD}} \cite{pmd}\textbf{:} A widely adopted static analyzer that flags potential inefficiencies and bad practices in Java code through rigorous rule matching. In this study, we specifically task \textsc{PMD} with identifying \textit{DC} instances.
\end{itemize}
These tools serve as foundational reference points in our study and are extensively cited in prior literature \cite{sharma2021qscored, shah2023mining, barbosa2020revealing, fernandes2016review, aniche2018code, wu2024ismell}. Furthermore, since these tools were utilized in our initial code smell collection phase (see Section \ref{subsec: Code Smell Detection}), comparing against them provides critical insight into how much improvement our PEFT-based approach offers over the original labeling sources.
}

\subsubsection{\textcolor{black}{DL-based Approaches}}
\textcolor{black}{We incorporate two representative DL models as baselines to evaluate the performance of established neural architectures:
\begin{itemize}
    \item \textbf{\textsc{DeepSmells}} \cite{ho2023fusion}\textbf{:} A model that utilizes a 1D-Convolutional Neural Network (1D-CNN) followed by a Long Short-Term Memory (LSTM) network to learn representations from tokenized source code sequences.
    \item \textbf{\textsc{AE-Dense}} \cite{sharma2021code}\textbf{:} An Autoencoder-based model designed to learn dense vector representations of code instances in an unsupervised manner.
\end{itemize}
Collectively, these comparative DL models represent the state-of-the-art in leveraging sequential patterns and latent feature learning for code smell detection, respectively.}

\subsubsection{\textcolor{black}{LLMs with In-Context Learning (ICL)}}
\textcolor{black}{To benchmark the performance of our fine-tuned models against the ``prompting'' paradigm without parameter updates for code smell detection, we leverage two powerful general-purpose LLMs. These models exemplify the latest advances in commercial and open-weight architectures under the ICL paradigm:
\begin{itemize}
    \item \textbf{GPT-4o-mini} \cite{hurst2024gpt4o}\textbf{:} One of the most advanced proprietary LLMs developed by OpenAI, known for its superior reasoning capabilities and code understanding in lightweight deployments.
    \item \textbf{DeepSeek-v3} \cite{liu2024deepseek}\textbf{:} A current cutting-edge open-source mixture-of-experts (MoE) model (671B parameters), which serves as a strong reference for the performance of large-scale open-weight models.
\end{itemize}
Both models have demonstrated exceptional capabilities across natural language and programming tasks \cite{rasheed2024task, bruni2025benchmarking, liu2024deepseek, jahin2025unveiling}. We assess their effectiveness in code smell detection under three ICL settings: zero-shot, few-shot, and a criteria-informed setting that provides explicit labeling rules (refer to Section \ref{subsec:results of RQ4} for details).}

\subsection{Metrics}
We use metrics that align with previous research \cite{sharma2021code, ho2023fusion, pecorelli2019comparing} to maintain consistency in evaluating the effectiveness of PEFT methods for code smell detection on LLMs. Specifically, we utilize precision, recall, F1-score, and MCC (Matthews Correlation Coefficient) for assessing the performance of detecting \textit{CC} and \textit{CM}. We include MCC because, unlike other metrics like F1-score, MCC accounts for true negative instances \cite{yao2020assessing}.
These metrics are calculated using the following formulas:

\begin{equation}
    \text{Precision} = \frac{TP}{TP + FP}
\end{equation}
\begin{equation}
    \text{Recall} = \frac{TP}{TP + FN}
\end{equation}
\begin{equation}
    \text{F1-Score} = 2 \times \frac{\text{Precision} \times \text{Recall}}{\text{Precision} + \text{Recall}}
\end{equation}
\begin{equation}
    \text{MCC} = \frac{TP \times TN - FP \times FN}{\sqrt{(TP + FP)(TP + FN)(TN + FP)(TN + FN)}}
\end{equation}
where $TP$, $TN$, $FP$, $FN$ represent true positives, true negatives, false positives, and false negatives, respectively.

\subsection{Implementation Details}
\label{subsec:implementation details}
The code for our experiments is implemented by the PEFT library\footnote{\url{https://huggingface.co/docs/peft/index}} from Hugging Face. After reviewing prior studies \cite{liu2023empirical, liu2024delving, li2024comprehensive, wang2023one, weyssow2023exploring} and observing the results of pilot experiments, we set the learning rate to $3e^{-4}$ for all PEFT methods and $1e^{-5}$ for full fine-tuning the  SLMs. Throughout all the fine-tuning experiments, we utilized the AdamW optimizer \cite{loshchilov2017decoupled} and fixed the Transformer input window size at 1,024. The choice to limit the input length to 1,024 tokens was driven by hardware constraints. As the GPU memory consumption of transformer-based models grows roughly quadratically with input length \cite{tay2022efficient}, this setting enables all selected base models to train within available GPU memory capacity.
All the models were trained for 10 epochs, and we selected the epochs at which models exhibit minimum loss on the validation sets for inference. 

Table \ref{tab:hyperparameter setting} shows the hyper-parameter settings of different PEFT methods. 
\textcolor{black}{Due to resource constraints, the batch size was set to 1 for LLMs with large parameter sizes to avoid out-of-memory issues. To ensure consistency and comparability, we adopted a batch size of 1 for all experiments.}
We used a fixed random seed (42) across all runs to ensure experimental reproducibility. For prompt tuning and prefix tuning, we used 20 virtual tokens, which are learnable vectors prepended to the input and optimized during training. This setting is widely used in the literature and offers a balance between representational capacity and computational cost \cite{li2021prefix, weyssow2023exploring}. As for LoRA, we set the rank $r$ to 8, the scaling factor $\alpha$ to 16, and the dropout rate to 0.1 to prevent overfitting, following configurations reported in \cite{liu2024delving}. For both LoRA and ${IA}^3$, we applied parameter-efficient adaptation to the core components of the Transformer architecture, including the query ($Q$), key ($K$), and value ($V$) projections in the attention layers. In the case of $(\text{IA})^3$, we also included the Feed-Forward Network (FFN) for adaptation (see Section \ref{subsec:explainations of peft}).
We executed all models on a single Linux GPU server equipped with a 40GB NVIDIA A100 GPU. Additionally, we have released our datasets and source code online for replication purpose \cite{replicationPackage} where more details can be found.

\begin{table}[htbp]
\footnotesize
\centering
\caption{Hyper-parameter settings of different PEFT methods and full fine-tuning}
\label{tab:hyperparameter setting}
\begin{tabular}{m{1.8cm}<{\arraybackslash}|m{4cm}<{\centering\arraybackslash}|m{1.3cm}<{\centering\arraybackslash}|m{1.2cm}<{\centering\arraybackslash}}
\toprule
\textbf{PEFT Method}  & \textbf{Hyper-parameter Setting}               & \textbf{Batch Size} & \textbf{Seed} \\ \cmidrule{1-4}      
Prompt Tuning         & \multirow{2}{*}{Number of virtual tokens: 20} & \multirow{9}{*}{1}  & \multirow{9}{*}{42} \\ \cmidrule{1-1} 
Prefix Tuning         &                                               & & \\ \cmidrule{1-2}
\multirow{4}{*}{LoRA} & Rank $r$: 8                                   & & \\
                      & $alpha$: 16                                   & & \\
                      & $dropout$: 0.1                                & & \\
                      & Target modules: $Q$, $K$, and $V$             & & \\ \cmidrule{1-2}
$(IA)^3$              & Target modules: $Q$, $K$, $V$, and FFN        & & \\ \cmidrule{1-2}
\textcolor{black}{Full Fine-tuning}       & None                                & & \\
\bottomrule
\end{tabular}
\end{table}

\section{Results}
\label{sec:results}
\textcolor{black}{In this section, we present the results and findings for the four RQs formulated in Section \ref{subsec:research questions}. For each RQ, we first specify the particular experimental configuration and execution strategy based on the general setup in Section \ref{sec:experiment setup}, followed by a detailed analysis of the results.}

\subsection{RQ1: What is the effectiveness of different PEFT methods and full fine-tuning on various SLMs and LLMs?}
\label{sec:RQ1}
Since our objective is to demonstrate the effectiveness of PEFT methods on both small and large LMs for code smell detection, we conducted a series of experiments involving four PEFT methods (i.e., prompt tuning, prefix tuning, LoRA, and $(IA)^{3}$) and a group of models with different parameter sizes. Each PEFT method was employed to fine-tune the 9 base models (see Section~\ref{sec:base model selection}). Note that we only applied LoRA and $(IA)^3$ to Code-T5 as the current PEFT library does not support fine-tuning encoder-decoder models with prompt tuning and prefix tuning. We full fine-tuned all the small models due to resource constraints, as the computational demands of tuning all parameters in LLMs would surpass the available 40GB of GPU memory. 
Table \ref{tab:results_of_RQ1_CC_CM} and Table \ref{tab:results_of_RQ1_FE_DC} present the comprehensive comparison of small LMs and LLMs tuning with various PEFT techniques for detecting \textit{CC}, \textit{CM}, \textit{FE}, and \textit{DC}, respectively.

\begin{table}[htbp]
\centering
\footnotesize
\caption{\textcolor{black}{Results of small LMs and LLMs with different PEFT methods on detecting \textit{CC} and \textit{CM} (\textbf{bold}: highest values per model, \underline{\textbf{underline}}: highest values among all models)}}
\label{tab:results_of_RQ1_CC_CM}
\begin{tabular}{m{0.8cm}<{\centering\arraybackslash}m{2.5cm}<{\arraybackslash}m{1.7cm}<{\arraybackslash}|m{1.1cm}<{\centering\arraybackslash}|m{1.2cm}<{\centering\arraybackslash}|m{1.2cm}<{\centering\arraybackslash}|m{1.2cm}<{\centering\arraybackslash}|m{1.2cm}<{\centering\arraybackslash}}
\toprule
& \multirow{2}{*}{\textbf{Model}} & \multirow{2}{*}{\textbf{Method}} & \textcolor{black}{\multirow{2}{*}{\textbf{\#Params}}} & \multicolumn{2}{c|}{\textbf{CC}} & \multicolumn{2}{c}{\textbf{CM}} \\ \cmidrule{5-8}
 &  &  &  & \textbf{F1} & \textbf{MCC} & \textbf{F1} & \textbf{MCC} \\ \cmidrule{1-8}
\multirow{19}{*}{\textit{SLMs}}
& \multirow{5}{*}{CodeBERT} & Full FT       & \textcolor{black}{125M} & \textcolor{gray}{70.54\%} & \textcolor{gray}{44.41\%} & \textcolor{gray}{80.57\%} & \textcolor{gray}{61.19\%} \\
&                           & Prompt Tuning & \textcolor{black}{0.61M} & \textcolor{gray}{62.38\%} & \textcolor{gray}{25.90\%} & \textcolor{gray}{76.13\%} & \textcolor{gray}{54.00\%} \\
&                           & Prefix Tuning & \textcolor{black}{0.96M} & \textcolor{gray}{63.88\%} & \textcolor{gray}{27.87\%} & \textbf{81.33\%} & \textcolor{gray}{62.94\%} \\
&                           & LoRA          & \textcolor{black}{1.03M} & \textbf{73.04\%} & \textbf{46.10\%} & \textcolor{gray}{71.00\%} & \textcolor{gray}{49.79\%} \\
&                           & $(IA)^3$      & \textcolor{black}{0.67M} & \textcolor{gray}{69.44\%} & \textcolor{gray}{40.49\%} & \textcolor{gray}{81.32\%} & \textbf{62.96\%} \\ \cmidrule{2-8}
& \multirow{5}{*}{GraphCodeBERT} & Full FT       & \textcolor{black}{125M} & \textbf{74.01\%} & \textbf{48.08\%} & \textcolor{gray}{80.66\%} & \textbf{61.33\%} \\
&                                & Prompt Tuning & \textcolor{black}{0.61M} & \textcolor{gray}{65.24\%} & \textcolor{gray}{30.81\%} & \textcolor{gray}{78.90\%} & \textcolor{gray}{58.60\%} \\
&                                & Prefix Tuning & \textcolor{black}{0.96M} & \textcolor{gray}{69.82\%} & \textcolor{gray}{39.97\%} & \textcolor{gray}{79.18\%} & \textcolor{gray}{60.46\%} \\
&                                & LoRA          & \textcolor{black}{1.03M} & \textcolor{gray}{73.15\%} & \textcolor{gray}{46.30\%} & \textbf{80.03\%} & \textcolor{gray}{60.93\%} \\
&                                & $(IA)^3$      & \textcolor{black}{0.67M} & \textcolor{gray}{72.50\%} & \textcolor{gray}{45.00\%} & \textcolor{gray}{80.00\%} & \textcolor{gray}{60.45\%} \\ \cmidrule{2-8}
& \multirow{5}{*}{UnixCoder} & Full FT       & \textcolor{black}{125M} & \textcolor{gray}{73.04\%} & \textcolor{gray}{46.10\%} & \textcolor{gray}{82.39\%} & \textcolor{gray}{64.89\%} \\
&                            & Prompt Tuning & \textcolor{black}{0.61M} & \textcolor{gray}{69.35\%} & \textcolor{gray}{38.70\%} & \textcolor{gray}{78.66\%} & \textcolor{gray}{57.95\%} \\
&                            & Prefix Tuning & \textcolor{black}{0.96M} & \textcolor{gray}{71.63\%} & \textcolor{gray}{43.26\%} & \textbf{82.71\%} & \textbf{65.84\%} \\
&                            & LoRA          & \textcolor{black}{1.03M} & \textcolor{gray}{72.19\%} & \textcolor{gray}{46.04\%} & \textcolor{gray}{73.18\%} & \textcolor{gray}{51.52\%} \\
&                            & $(IA)^3$      & \textcolor{black}{0.67M} & \textbf{73.47\%} & \textbf{46.98\%} & \textcolor{gray}{82.23\%} & \textcolor{gray}{64.47\%} \\ \cmidrule{2-8}
& \multirow{3}{*}{CodeT5} & Full FT  & \textcolor{black}{220M} & \textbf{73.01\%} & \textbf{46.19\%} & \textcolor{gray}{81.13\%} & \textbf{62.74\%} \\
&                         & LoRA     & \textcolor{black}{1.33M} & \textcolor{gray}{33.33\%} & \textcolor{gray}{0.00\%} & \textbf{83.29\%} & \textcolor{gray}{-2.95\%} \\
&                         & $(IA)^3$ & \textcolor{black}{0.16M} & \textcolor{gray}{43.38\%} & \textcolor{gray}{-4.05\%} & \textcolor{gray}{40.32\%} & \textcolor{gray}{16.70\%} \\
\midrule
\midrule
\multirow{22}{*}{\textit{LLMs}}
& \multirow{4}{*}{StarCoderBase-1B} & Prompt Tuning & \textcolor{black}{0.05M} & \textcolor{gray}{71.07\%} & \textcolor{gray}{42.21\%} & \textcolor{gray}{80.05\%} & \textcolor{gray}{60.11\%} \\
&                                   & Prefix Tuning & \textcolor{black}{1.97M} & \textcolor{gray}{70.10\%} & \textcolor{gray}{40.23\%} & \textcolor{gray}{76.88\%} & \textcolor{gray}{54.01\%} \\
&                                   & LoRA          & \textcolor{black}{3.59M} & \textcolor{gray}{41.41\%} & \textcolor{gray}{-1.62\%} & \textcolor{gray}{45.41\%} & \textcolor{gray}{7.01\%} \\
&                                   & $(IA)^3$      & \textcolor{black}{0.31M} & \textbf{74.27\%} & \textbf{49.01\%} & \textbf{80.90\%} & \textbf{62.01\%} \\ \cmidrule{2-8}
& \multirow{4}{*}{StarCoderBase-3B} & Prompt Tuning & \textcolor{black}{0.06M} & \textcolor{gray}{69.12\%} & \textcolor{gray}{38.29\%} & \textcolor{gray}{81.87\%} & \textcolor{gray}{63.85\%} \\
&                                   & Prefix Tuning & \textcolor{black}{4.06M} & \textcolor{gray}{67.81\%} & \textcolor{gray}{35.69\%} & \textcolor{gray}{78.25\%} & \textcolor{gray}{56.92\%} \\
&                                   & LoRA          & \textcolor{black}{7.38M} & \textcolor{gray}{40.13\%} & \textcolor{gray}{7.93\%} & \textcolor{gray}{68.41\%} & \textcolor{gray}{46.00\%} \\
&                                   & $(IA)^3$      & \textcolor{black}{0.62M} & \underline{\textbf{74.68\%}} & \underline{\textbf{49.96\%}} & \textbf{82.07\%} & \textbf{64.81\%} \\ \cmidrule{2-8}
& \multirow{4}{*}{Deepseek-Coder-1.3B} & Prompt Tuning & \textcolor{black}{0.05M} & \textcolor{gray}{70.21\%} & \textcolor{gray}{40.44\%} & \textcolor{gray}{81.77\%} & \textcolor{gray}{63.75\%} \\
&                                      & Prefix Tuning & \textcolor{black}{1.97M} & \textcolor{gray}{69.38\%} & \textcolor{gray}{39.62\%} & \textcolor{gray}{79.06\%} & \textcolor{gray}{58.36\%} \\
&                                      & LoRA          & \textcolor{black}{2.36M} & \textcolor{gray}{69.44\%} & \textcolor{gray}{40.49\%} & \textcolor{gray}{78.42\%} & \textcolor{gray}{57.96\%} \\
&                                      & $(IA)^3$      & \textcolor{black}{0.28M} & \textbf{72.50\%} & \textbf{45.01\%} & \textbf{81.87\%} & \textbf{63.82\%} \\ \cmidrule{2-8}
& \multirow{4}{*}{Deepseek-Coder-6.7B} & Prompt Tuning & \textcolor{black}{0.09M} & \textcolor{gray}{67.93\%} & \textcolor{gray}{35.87\%} & \textcolor{gray}{81.19\%} & \textcolor{gray}{63.11\%} \\
&                                      & Prefix Tuning & \textcolor{black}{5.25M} & \textcolor{gray}{68.80\%} & \textcolor{gray}{37.62\%} & \textcolor{gray}{80.09\%} & \textcolor{gray}{60.57\%} \\
&                                      & LoRA          & \textcolor{black}{6.30M} & \textcolor{gray}{69.98\%} & \textcolor{gray}{40.05\%} & \textcolor{gray}{81.34\%} & \textcolor{gray}{63.63\%} \\
&                                     & $(IA)^3$       & \textcolor{black}{0.75M} & \textbf{70.84\%} & \textbf{42.36\%} & \underline{\textbf{83.34\%}} & \underline{\textbf{66.88\%}} \\ \cmidrule{2-8}
& \multirow{4}{*}{CodeLlama-7B} & Prompt Tuning & \textcolor{black}{0.09M} & \textcolor{gray}{68.68\%} & \textcolor{gray}{37.92\%} & \textcolor{gray}{81.25\%} & \textcolor{gray}{63.43\%} \\
&                               & Prefix Tuning & \textcolor{black}{5.25M} & \textcolor{gray}{53.41\%} & \textcolor{gray}{15.95\%} & \textcolor{gray}{80.77\%} & \textcolor{gray}{62.08\%} \\
&                               & LoRA          & \textcolor{black}{6.30M} & \textcolor{gray}{70.12\%} & \textcolor{gray}{40.70\%} & \textcolor{gray}{80.70\%} & \textcolor{gray}{62.59\%} \\
&                               & $(IA)^3$      & \textcolor{black}{0.75M} & \textbf{70.48\%} & \textbf{42.32\%} & \textbf{83.00\%} & \textbf{66.17\%} \\
\bottomrule
\end{tabular}
\end{table}
\begin{table}[htbp]
\centering
\footnotesize
\caption{\textcolor{black}{Results of small LMs and LLMs with different PEFT methods on detecting \textit{FE} and \textit{DC} (\textbf{bold}: highest values per model, \underline{\textbf{underline}}: highest values among all models)}}
\label{tab:results_of_RQ1_FE_DC}
\begin{tabular}{m{0.8cm}<{\centering\arraybackslash}m{2.5cm}<{\arraybackslash}m{1.7cm}<{\arraybackslash}|m{1.1cm}<{\centering\arraybackslash}|m{1.2cm}<{\centering\arraybackslash}|m{1.2cm}<{\centering\arraybackslash}|m{1.2cm}<{\centering\arraybackslash}|m{1.2cm}<{\centering\arraybackslash}}
\toprule
& \multirow{2}{*}{\textbf{Model}} & \multirow{2}{*}{\textbf{Method}} & \textcolor{black}{\multirow{2}{*}{\textbf{\#Params}}} & \multicolumn{2}{c|}{\textbf{FE}} & \multicolumn{2}{c}{\textbf{DC}} \\ \cmidrule{5-8}
 &  &  &  & \textbf{F1} & \textbf{MCC} & \textbf{F1} & \textbf{MCC} \\ \cmidrule{1-8}
\multirow{19}{*}{\textit{SLMs}}
& \multirow{5}{*}{CodeBERT} & Full FT       & \textcolor{black}{125M} & \textbf{74.09\%} & \textbf{48.80\%} & \textcolor{gray}{81.98\%} & \textcolor{gray}{65.60\%}\\
&                           & Prompt Tuning & \textcolor{black}{0.61M} & \textcolor{gray}{56.17\%} & \textcolor{gray}{13.07\%} & \textcolor{gray}{81.56\%} & \textcolor{gray}{64.57\%} \\
&                           & Prefix Tuning & \textcolor{black}{0.96M} & \textcolor{gray}{62.82\%} & \textcolor{gray}{25.93\%} & \textcolor{gray}{81.81\%} & \textcolor{gray}{65.43\%} \\
&                           & LoRA          & \textcolor{black}{1.03M} & \textcolor{gray}{67.44\%} & \textcolor{gray}{36.17\%} & \textbf{83.17\%} & \textbf{68.48\%} \\
&                           & $(IA)^3$      & \textcolor{black}{0.67M} & \textcolor{gray}{66.49\%} & \textcolor{gray}{38.48\%} & \textcolor{gray}{83.30\%} & \textcolor{gray}{68.18\%} \\ \cmidrule{2-8}
& \multirow{5}{*}{GraphCodeBERT} & Full FT       & \textcolor{black}{125M} & \textbf{77.40\%} & \textbf{54.95\%} & \textcolor{gray}{82.66\%} & \textcolor{gray}{67.42\%} \\
&                                & Prompt Tuning & \textcolor{black}{0.61M} & \textcolor{gray}{66.88\%} & \textcolor{gray}{37.49\%} & \textcolor{gray}{83.75\%} & \textcolor{gray}{68.95\%} \\
&                                & Prefix Tuning & \textcolor{black}{0.96M} & \textcolor{gray}{74.19\%} & \textcolor{gray}{48.39\%} & \underline{\textbf{89.26\%}} & \underline{\textbf{79.26\%}} \\
&                                & LoRA          & \textcolor{black}{1.03M} & \textcolor{gray}{72.23\%} & \textcolor{gray}{46.36\%} & \textcolor{gray}{85.85\%} & \textcolor{gray}{72.40\%} \\
&                                & $(IA)^3$      & \textcolor{black}{0.67M} & \textcolor{gray}{69.84\%} & \textcolor{gray}{45.48\%} & \textcolor{gray}{85.58\%} & \textcolor{gray}{72.56\%} \\ \cmidrule{2-8}
& \multirow{5}{*}{UnixCoder} & Full FT       & \textcolor{black}{125M} & \textcolor{gray}{77.04\%} & \textbf{56.76\%} & \textcolor{gray}{83.51\%} & \textbf{69.49\%} \\
&                            & Prompt Tuning & \textcolor{black}{0.61M} & \textbf{77.32\%} & \textcolor{gray}{55.30\%} & \textcolor{gray}{83.68\%} & \textcolor{gray}{68.74\%} \\
&                            & Prefix Tuning & \textcolor{black}{0.96M} & \textcolor{gray}{75.80\%} & \textcolor{gray}{51.64\%} & \textcolor{gray}{83.89\%} & \textcolor{gray}{69.28\%} \\
&                            & LoRA          & \textcolor{black}{1.03M} & \textcolor{gray}{70.97\%} & \textcolor{gray}{41.94\%} & \textcolor{gray}{83.60\%} & \textcolor{gray}{68.65\%} \\
&                            & $(IA)^3$      & \textcolor{black}{0.67M} & \textcolor{gray}{73.95\%} & \textcolor{gray}{49.32\%} & \textbf{83.96\%} & \textcolor{gray}{69.44\%} \\ \cmidrule{2-8}
& \multirow{3}{*}{CodeT5} & Full FT  & \textcolor{black}{220M} & \textbf{66.49\%} & \textbf{38.48\%} & \textbf{83.46\%} & \textbf{69.12\%} \\
&                         & LoRA     & \textcolor{black}{1.33M} & \textcolor{gray}{54.07\%} & \textcolor{gray}{10.02\%} & \textcolor{gray}{70.03\%} & \textcolor{gray}{49.06\%} \\
&                         & $(IA)^3$ & \textcolor{black}{0.16M} & \textcolor{gray}{54.65\%} & \textcolor{gray}{9.76\%} & \textcolor{gray}{46.79\%} & \textcolor{gray}{-2.66\%} \\
\midrule
\midrule
\multirow{22}{*}{\textit{LLMs}}
& \multirow{4}{*}{StarCoderBase-1B} & Prompt Tuning & \textcolor{black}{0.05M} & \textcolor{gray}{72.23\%} & \textcolor{gray}{46.36\%} & \textcolor{gray}{84.66\%} & \textcolor{gray}{71.21\%} \\
&                                   & Prefix Tuning & \textcolor{black}{1.97M} & \textcolor{gray}{75.65\%} & \textcolor{gray}{52.30\%} & \textbf{85.44\%} & \textbf{72.19\%} \\
&                                   & LoRA          & \textcolor{black}{3.59M} & \textbf{78.58\%} & \textbf{60.68\%} & \textcolor{gray}{79.30\%} & \textcolor{gray}{58.85\%} \\
&                                   & $(IA)^3$      & \textcolor{black}{0.31M} & \textcolor{gray}{73.95\%} & \textcolor{gray}{49.32\%} & \textcolor{gray}{84.96\%} & \textcolor{gray}{71.71\%} \\ \cmidrule{2-8}
& \multirow{4}{*}{StarCoderBase-3B} & Prompt Tuning & \textcolor{black}{0.06M} & \textcolor{gray}{68.96\%} & \textcolor{gray}{39.74\%} & \textcolor{gray}{83.63\%} & \textcolor{gray}{68.47\%} \\
&                                   & Prefix Tuning & \textcolor{black}{4.06M} & \textcolor{gray}{67.61\%} & \textcolor{gray}{35.78\%} & \textbf{86.56\%} & \textbf{74.06\%} \\
&                                   & LoRA          & \textcolor{black}{7.38M} & \textbf{70.85\%} & \textbf{42.29\%} & \textcolor{gray}{83.70\%} & \textcolor{gray}{68.64\%} \\
&                                   & $(IA)^3$      & \textcolor{black}{0.62M} & \textcolor{gray}{69.35\%} & \textcolor{gray}{38.73\%} & \textcolor{gray}{84.90\%} & \textcolor{gray}{71.46\%} \\ \cmidrule{2-8}
& \multirow{4}{*}{DeepSeek-Coder-1.3B} & Prompt Tuning & \textcolor{black}{0.05M} & \textcolor{gray}{78.98\%} & \textcolor{gray}{58.34\%} & \textcolor{gray}{84.17\%} & \textcolor{gray}{69.99\%} \\
&                                      & Prefix Tuning & \textcolor{black}{1.97M} & \textcolor{gray}{66.88\%} & \textcolor{gray}{37.49\%} & \textcolor{gray}{82.44\%} & \textcolor{gray}{64.90\%} \\
&                                      & LoRA          & \textcolor{black}{2.36M} & \textcolor{gray}{78.98\%} & \textcolor{gray}{58.34\%} & \textcolor{gray}{82.72\%} & \textcolor{gray}{67.68\%} \\
&                                      & $(IA)^3$      & \textcolor{black}{0.28M} & \textbf{\underline{83.87\%}} & \textbf{\underline{67.74\%}} & \textbf{85.84\%} & \textbf{72.57\%} \\ \cmidrule{2-8}
& \multirow{4}{*}{DeepSeek-Coder-6.7B} & Prompt Tuning & \textcolor{black}{0.09M} & \textcolor{gray}{78.89\%} & \textcolor{gray}{58.83\%} & \textbf{85.56\%} & \textbf{72.72\%} \\
&                                      & Prefix Tuning & \textcolor{black}{5.25M} & \textbf{80.56\%} & \textbf{61.81\%} & \textcolor{gray}{83.03\%} & \textcolor{gray}{68.14\%} \\
&                                      & LoRA          & \textcolor{black}{6.30M} & \textcolor{gray}{77.32\%} & \textcolor{gray}{55.30\%} & \textcolor{gray}{83.43\%} & \textcolor{gray}{68.64\%} \\
&                                      & $(IA)^3$      & \textcolor{black}{0.75M} & \textcolor{gray}{77.32\%} & \textcolor{gray}{55.30\%} & \textcolor{gray}{85.34\%} & \textcolor{gray}{72.34\%} \\ \cmidrule{2-8}
& \multirow{4}{*}{CodeLlama-7B} & Prompt Tuning & \textcolor{black}{0.09M} & \textcolor{gray}{72.52\%} & \textcolor{gray}{45.37\%} & \textbf{85.49\%} & \textbf{72.55\%} \\
&                               & Prefix Tuning & \textcolor{black}{5.25M} & \textcolor{gray}{63.57\%} & \textcolor{gray}{30.67\%} & \textcolor{gray}{82.30\%} & \textcolor{gray}{64.61\%} \\
&                               & LoRA          & \textcolor{black}{6.30M} & \textbf{77.32\%} & \textbf{55.30\%} & \textcolor{gray}{83.10\%} & \textcolor{gray}{68.31\%} \\
&                               & $(IA)^3$      & \textcolor{black}{0.75M} & \textcolor{gray}{74.19\%} & \textcolor{gray}{48.39\%} & \textcolor{gray}{83.54\%} & \textcolor{gray}{68.45\%} \\
\bottomrule
\end{tabular}
\end{table}

\subsubsection{SLMs}
Overall, GraphCodeBERT proved the most effective in identifying \textit{CC} and \textit{DC}, while UnixCoder excelled at detecting \textit{CM} and \textit{FE}. In comparison, CodeT5 fine-tuned with PEFT methods performed the worst across all types of code smells. CodeBERT performed similarly to GraphCodeBERT, albeit slightly worse. For example, in \textit{CC} detection, whether using full fine-tuning or PEFT methods, CodeBERT consistently achieved lower F1-scores and MCC values compared to GraphCodeBERT, with differences ranging from 0.11\% to 5.93\% and 0.21\% to 12.10\%, respectively. This is reasonable because GraphCodeBERT is built upon CodeBERT and further incorporates code structure information (i.e., data flow) during the pre-training phase \cite{guo2021graphcodebert}. Across all SLMs, CodeT5 with PEFT methods significantly lags behind the other three models with PEFT methods in the code smell detection task, which requires both method- and class-level code understanding. When fine-tuned with PEFT methods, the MCC values of CodeBERT even became negative, reaching -4.05\% for LoRA on \textit{CC} and -2.66\% for $(IA)^3$ on \textit{DC}. This suggests that CodeT5 with PEFT methods struggles to perform effectively in code smell detection, highlighting the limitations of its PEFT adaptation. In contrast, the fully fine-tuned CodeT5 demonstrated acceptable performance in code smell detection. The stark performance gap between CodeT5 with full fine-tuning and PEFT methods could stem from CodeT5’s architecture, which may depend more on full fine-tuning for optimal adaptation to the method-level code smell detection task. Compared to full fine-tuning, PEFT methods like LoRA and $(IA)^3$ might not provide sufficient adjustments to effectively adapt CodeT5’s encoder-decoder architecture.

\subsubsection{LLMs}
\label{sec:LLMs}
DeepSeek-Coder-1.3B stands out as the top-performing LLM, consistently achieving the highest or on-par F1-scores and MCC values across all fine-tuning methods when compared to other LLMs. 
Although the StarCoderBase models perform similarly to the DeepSeek-Coder models in detecting \textit{FE} and \textit{DC}, which involve class-level code understanding, their results are notably less stable for detecting \textit{CC} and \textit{CM}, which require method-level understanding. In particular, the performance of StarCoderBase models fine-tuned with LoRA degrades significantly on the detection of \textit{CC} and \textit{CM}.
Notably, when fine-tuned with PEFT techniques, models with larger parameter sizes within the same family do not surpass, and in some cases, even slightly underperform their smaller counterparts. For example, in \textit{FE} detection, StarCoderBase-1B and DeepSeek-Coder-1.3B generally achieve higher F1-scores and MCC values than StarCoderBase-3B and DeepSeek-Coder-6.7B.

\subsubsection{SLMs vs. LLMs}
\label{sec:Small LMs vs.LLMs}
For the detection of \textit{CC} and \textit{CM}, SLMs and LLMs demonstrate generally comparable performance. Notably, in the case of \textit{DC}, GraphCodeBERT, a representative SLM, significantly outperforms all other models, including LLMs. These results suggest that smaller models can remain competitive, or even superior, for certain types of code smells. In contrast, for \textit{FE}, LLMs fine-tuned with PEFT methods substantially outperform SLMs, likely due to the more complex and semantic nature of this smell type, which benefits from the richer contextual representations and broader knowledge captured by larger models.

\begin{figure}
    \centering
    \includegraphics[width=0.83\linewidth]{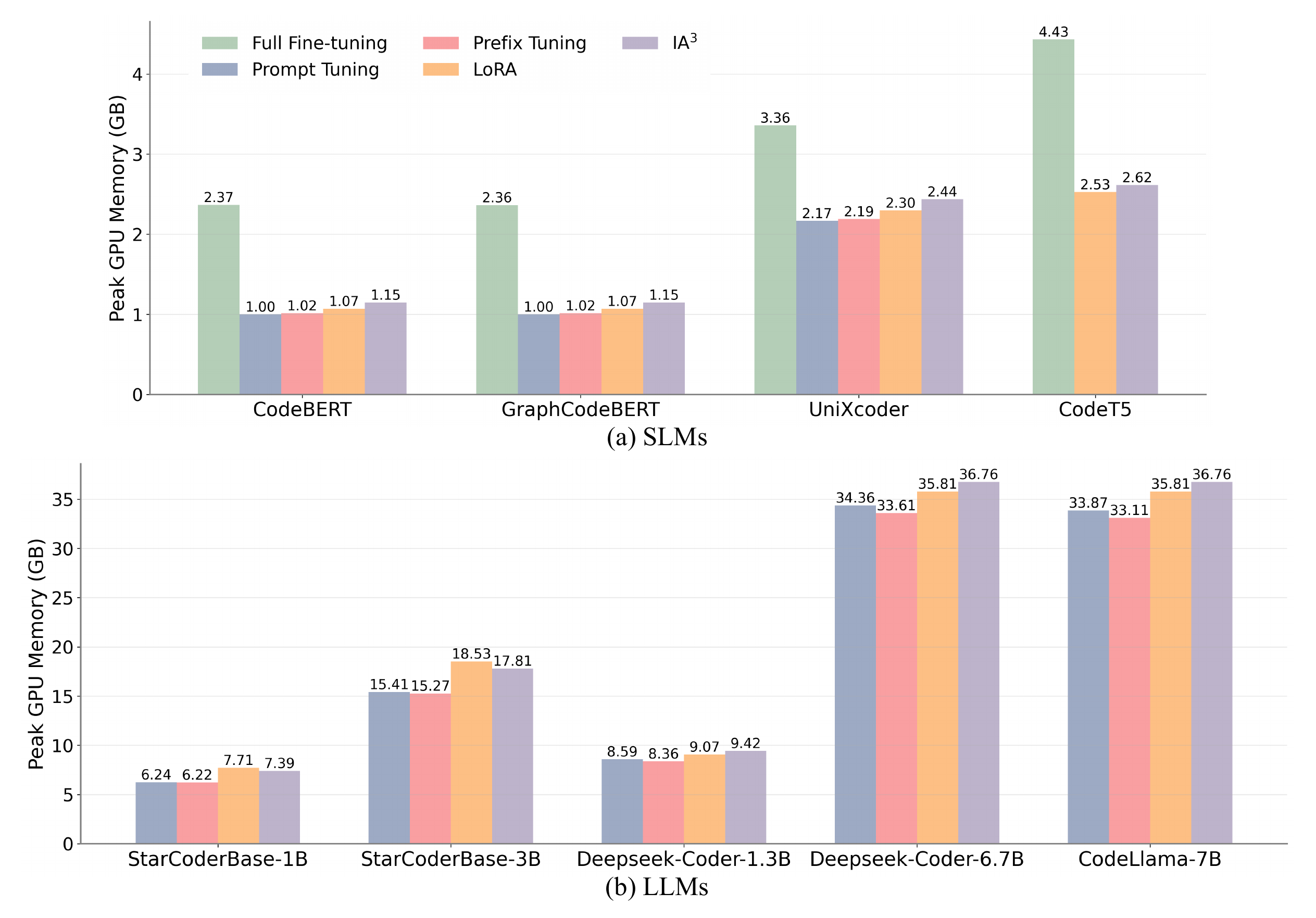}
    \caption{Peak GPU memory usage for detecting \textit{CC}: (a) SLMs with PEFT methods and Full Fine-Tuning; (b) LLMs with PEFT methods}
    \label{fig:peak gpu}
\end{figure}

Additionally, Figure \ref{fig:peak gpu} summarizes the peak GPU memory usage of both SLMs and LLMs with different PEFT methods on \textit{CC} detection. Comparing Figure \ref{fig:peak gpu} (a) and (b), we can observe that PEFT methods impose substantially higher peak GPU memory requirements on LLMs compared to SLMs, particularly for larger-sized LLMs.

\subsubsection{Method-level vs. Class-level Code Understanding}
\label{subsubsec:code understanding}
The detection of \textit{CC} and \textit{CM} requires method-level code understanding, whereas identifying \textit{FE} and \textit{DC} necessitates class-level analysis (see Section \ref{subsubsec:selected code smells}). By comparing the detection results of \textit{CC} and \textit{CM} with those of \textit{FE} and \textit{DC}, we observe that LMs demonstrate comparable effectiveness at both levels of code understanding. Notably, the models even achieve better performance on class-level code smells, as evidenced by the fact that the best-performing model yields significantly lower results on \textit{CC} compared to the other smell types.

\subsubsection{Best PEFT Method}
\label{subsubsec:Best PEFT method}
\textcolor{black}{Generally, for SLMs, both $(IA)^3$ and prefix tuning achieve strong performance across different models, while for LLMs, $(IA)^3$ consistently delivers the best results across all four types of code smells. To be specific, the optimal PEFT method for SLMs on the task of code smell detection depends on the specific model: for instance, CodeBERT performs best with $(IA)^3$, whereas UnixCoder tends to favor prefix tuning. In contrast, $(IA)^3$ outperforms other PEFT techniques for all LLMs studied, including StarCoderBase Models, Deepseek-Coder Models, and CodeLlama-7B. Additionally, prompt tuning can serve as an alternative for LLMs, though its effectiveness is generally lower. Notably, despite achieving promising results on certain models and code smell types, LoRA exhibits instability in performance and can yield markedly poor effectiveness in some cases. For example, on StarCoderBase-1B for \textit{CM} detection, LoRA’s MCC drops to 7.01\%, while the other three PEFT methods achieve a minimum MCC of 54.01\%, resulting in a disparity of over 40\%.}

\textcolor{black}{As shown in Tables \ref{tab:results_of_RQ1_CC_CM} and \ref{tab:results_of_RQ1_FE_DC}, the additional parameters injected by all PEFT methods are negligible relative to the total model size.
Prefix tuning and LoRA generally introduce more trainable parameters than prompt tuning and $(IA)^3$, with LoRA adding the most and prompt tuning the least. For SLMs, the absolute differences in the number of added parameters remain modest due to their limited model size, whereas for LLMs the gaps become substantially larger.}

\textcolor{black}{Overall, $(IA)^3$ introduces a relatively small amount of trainable parameters while delivering strong performance gains for both SLMs and LLMs in code smell detection. For SLMs, prefix tuning still remains a practical alternative; however, for LLMs it brings a considerably larger parameter overhead without providing comparable performance benefits.}

\subsubsection{PEFT Methods vs. Full Fine-tuning}
\label{sec:PEFT methods vs. Full Fine-tuning}
According to Tables \ref{tab:results_of_RQ1_CC_CM} and \ref{tab:results_of_RQ1_FE_DC}, all four PEFT methods achieve better or at least comparable effectiveness than full fine-tuning on most small models for code smell detection. 
For SLMs (excluding CodeT5), prompt tuning, prefix tuning, LoRA, and $(IA)^3$ generally achieve varying degrees of improvement in F1-scores and MCC values compared to full fine-tuning across most code smell types. However, for \textit{FE}, all PEFT methods perform significantly worse than full fine-tuning. 

\textcolor{black}{The ``\#Params'' columns in Tables \ref{tab:results_of_RQ1_CC_CM} and \ref{tab:results_of_RQ1_FE_DC} further show that all PEFT methods introduce a relatively small number of trainable parameters compared to full fine-tuning, typically ranging from tens to hundreds of millions. This demonstrates that, in most cases, PEFT approaches can match or even surpass the performance of standard full fine-tuning for code smell detection while updating far fewer parameters, highlighting their efficiency for the task of code smell detection.}

\textcolor{black}{In addition to reducing trainable parameters, PEFT methods also lower computational overhead, particularly peak GPU memory usage, without sacrificing performance.
Taking the detection of \textit{CC} as an example, Figure \ref{fig:peak gpu}(a) depicts the peak GPU memory consumption during the fine-tuning of the four SLMs. 
It can be observed that, juxtaposed with full fine-tuning, all PEFT methods lead to a substantial reduction in peak GPU memory usage, which further validates the efficiency of PEFT methods for code smell detection relative to full fine-tuning.}

\noindent\begin{center}
		\begin{tcolorbox}[colback=black!5, colframe=black!20, width=1.0\linewidth, arc=1mm, auto outer arc, boxrule=1.5pt]                       
            \textbf{Answer to RQ1:} Overall, LLMs perform slightly better than SLMs in code smell detection, except for \textit{DC}. Among the PEFT methods, both $(IA)^3$ and prefix tuning are highly effective for SLMs, whereas $(IA)^3$ consistently delivers the best results for LLMs.
            \textcolor{black}{All PEFT techniques either exceed or match full fine-tuning while requiring far fewer trainable parameters and significantly reducing peak GPU memory usage, underscoring their efficiency for code smell detection.}
		\end{tcolorbox}
\end{center}

\subsection{RQ2: What is the effectiveness of PEFT methods with different hyper-parameters?}
\begin{figure}
    \centering
    \includegraphics[width=0.9\linewidth]{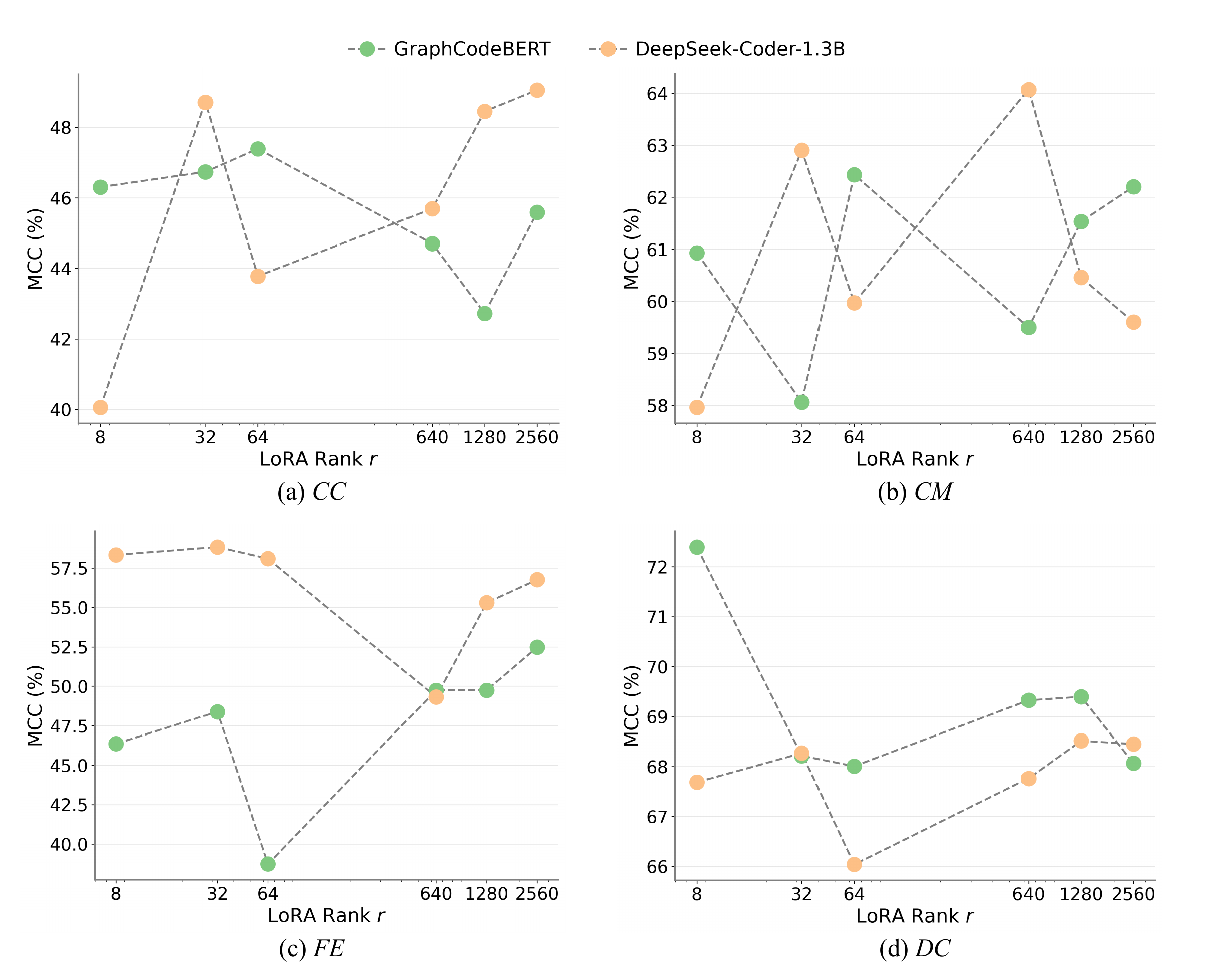}
    \caption{MCC results of GraphCodeBERT and DeepSeek-Coder-1.3B using LoRA for code smell detection under different rank settings: (a) \textit{CC}; (b) \textit{CM}; (c) \textit{FE}; (d) \textit{DC}}
    \label{fig:results of RQ2}
\end{figure}

Given the restrictions of our computational resources and popularity of LoRA, we focused on a single PEFT method, LoRA, to investigate how SLMs and LLMs perform under different hyper-parameter settings. Specifically, we used LoRA to tune a SLM (GraphCodeBERT) and an LLM (DeepSeek-Coder-1.3B), adapting the value of LoRA rank $r$ in the range of \{8, 32, 64, 640, 1,280, 2,560\}.  
At the same time, we kept other hyper-parameters of LoRA constant, with the LoRA alpha set to 16 and the LoRA dropout set to 0.1. 
The experiment results are presented in Figure \ref{fig:results of RQ2}.

\subsubsection{Results of RQ2}
The results in Figure \ref{fig:results of RQ2} show that for both GraphCodeBERT and DeepSeek-Coder-1.3B, the optimal $r$ value varies across different code smell types, indicating that the effectiveness of LMs on code smell detection is sensitive to the choice of LoRA configuration. Since the LoRA $r$ value directly controls the number of trainable parameters, this observation implies that different types of code smells may require different levels of model adaptability.

Furthermore, increasing the LoRA $r$ value from 8 to 2,560, which corresponds to an increase in the number of trainable parameters, does not uniformly lead to improved performance. Although a higher $r$ value increases the number of trainable parameters, which provides the model with greater capacity to learn, the MCC scores for detecting the four code smell types do not show a monotonically increasing trend. These findings indicate that expanding the number of trainable parameters alone does not guarantee better effectiveness, emphasizing the importance of careful tuning of the LoRA configuration rather than relying solely on scaling the model.

\noindent\begin{center}
		\begin{tcolorbox}[colback=black!5, colframe=black!20, width=1.0\linewidth, arc=1mm, auto outer arc, boxrule=1.5pt]                       
            {{\textbf{Answer to RQ2:} Increasing the number of trainable parameters by adjusting hyper-parameters does not necessarily ensure effectiveness enhancement for code smell detection task. Additionally, the effectiveness of LMs in detecting different code smells is sensitive to the configuration of PEFT methods (e.g., LoRA).}}
		\end{tcolorbox}
\end{center}

\subsection{RQ3: What is the effectiveness of PEFT methods in low-resource scenarios?}
\label{sec:RQ3}
To simulate low-resource scenarios, following previous studies \cite{wang2023one, liu2023empirical}, we randomly selected 50, 100, 250, and 500 samples for \textit{CC}, \textit{CM}, and \textit{DC}, respectively, based on the size of our dataset. Note that each subset is balanced, containing an equal number of positive and negative samples. We excluded \textit{FE} from this RQ, as the total number of available \textit{FE} samples in our training set is already below 500 (see Section \ref{subsec:data split}), making it inherently a low-resource task.
The experiment results are shown in Table \ref{tab:Results of GrahphCodeBERT-Base and DeepSeek-Coder-1.3B-Base with different PEFT methods and full fine-tuning on code smell detection in low-resource scenarios}. 

\begin{table*}[htbp]
\footnotesize
\centering
\caption{Results of GrahphCodeBERT and DeepSeek-Coder-1.3B with different PEFT methods and full fine-tuning on detecting \textit{CC}, \textit{CM}, and \textit{DC} in low-resource scenarios (\textbf{bold}: highest values per scenario for a model,  \underline{\textbf{underline}}: highest values among all scenarios for a model)}
\label{tab:Results of GrahphCodeBERT-Base and DeepSeek-Coder-1.3B-Base with different PEFT methods and full fine-tuning on code smell detection in low-resource scenarios}
\begin{tabular}{m{2.5cm}<{\arraybackslash}m{2.4cm}<{\centering\arraybackslash}m{1.8cm}<{\arraybackslash}|m{1.2cm}<{\centering\arraybackslash}|m{1.2cm}<{\centering\arraybackslash}|m{1.2cm}<{\centering\arraybackslash}}
\toprule
\multicolumn{1}{c}{{\textbf{Model}}} & \textbf{\#Training Samples} & \textbf{Method} & \textbf{CC} & \textbf{CM} & \textbf{DC} \\ \cmidrule{1-6}
\multirow{21}{*}{GraphCodeBERT}
& \multirow{5}{*}{50} & Full FT       & \textbf{19.23\%} & \textbf{54.61\%} & \textbf{62.76\%} \\ 
&                     & Prompt Tuning & \textcolor{gray}{-0.90\%} & \textcolor{gray}{38.19\%} & \textcolor{gray}{59.51\%} \\
&                     & Prefix Tuning & \textcolor{gray}{6.47\%} & \textcolor{gray}{31.64\%} & \textcolor{gray}{50.79\%} \\
&                     & LoRA          & \textcolor{gray}{7.29\%} & \textcolor{gray}{44.99\%} & \textcolor{gray}{62.51\%} \\
&                     & $(IA)^3$      & \textcolor{gray}{11.84\%} & \textcolor{gray}{45.15\%} & \textcolor{gray}{61.40\%} \\ \cmidrule{2-6}
& \multirow{5}{*}{100} & Full FT       & \textbf{24.67\%} & \textbf{51.58\%} & \textcolor{gray}{68.28\%} \\
&                      & Prompt Tuning & \textcolor{gray}{3.91\%} & \textcolor{gray}{39.03\%} & \textcolor{gray}{61.49\%} \\
&                      & Prefix Tuning & \textcolor{gray}{15.32\%} & \textcolor{gray}{38.87\%} & \textcolor{gray}{54.19\%} \\
&                      & LoRA          & \textcolor{gray}{23.14\%} & \textcolor{gray}{46.98\%} & \textbf{68.71\%} \\
&                      & $(IA)^3$      & \textcolor{gray}{23.66\%} & \textcolor{gray}{43.99\%} & \textcolor{gray}{65.62\%} \\ \cmidrule{2-6}
& \multirow{5}{*}{250} & Full FT       & \textcolor{gray}{28.38\%} & \textbf{55.36\%} & \textcolor{gray}{64.93\%} \\
&                      & Prompt Tuning & \textcolor{gray}{13.05\%} & \textcolor{gray}{42.03\%} & \textcolor{gray}{64.69\%} \\
&                      & Prefix Tuning & \textcolor{gray}{20.02\%} & \textcolor{gray}{47.02\%} & \textcolor{gray}{62.19\%} \\
&                      & LoRA          & \textbf{31.11\%} & \textcolor{gray}{51.91\%} & \textbf{66.79\%} \\
&                      & $(IA)^3$      & \textcolor{gray}{28.49\%} & \textcolor{gray}{52.45\%} & \textcolor{gray}{64.05\%} \\ \cmidrule{2-6}
& \multirow{5}{*}{500} & Full FT       & \textcolor{gray}{34.85\%} & \textcolor{gray}{57.03\%} & \textcolor{gray}{67.64\%} \\
&                      & Prompt Tuning & \textcolor{gray}{18.87\%} & \textcolor{gray}{47.50\%} & \textcolor{gray}{63.41\%} \\
&                      & Prefix Tuning & \textcolor{gray}{25.87\%} & \textcolor{gray}{53.05\%} & \textcolor{gray}{64.63\%} \\
&                      & LoRA          & \underline{\textbf{36.52\%}} & \textbf{58.31\%} & \textcolor{gray}{67.83\%} \\
&                      & $(IA)^3$      & \textcolor{gray}{34.14\%} & \textcolor{gray}{56.22\%} & \textbf{68.58\%} \\ 
\midrule
\midrule
\multirow{21}{*}{DeepSeek-Coder-1.3B}
& \multirow{5}{*}{50} & Full FT       & \textbf{11.38\%} & \textbf{42.46\%} & \textcolor{gray}{68.08\%} \\
&                     & Prompt Tuning & \textcolor{gray}{5.88\%} & \textcolor{gray}{25.99\%} & \textcolor{gray}{48.34\%} \\
&                     & Prefix Tuning & \textcolor{gray}{-3.70\%} & \textcolor{gray}{43.16\%} & \textcolor{gray}{41.93\%} \\
&                     & LoRA          & \textcolor{gray}{-1.48\%} & \textcolor{gray}{36.74\%} & \textbf{71.84\%} \\
&                     & $(IA)^3$      & \textcolor{gray}{1.54\%} & \textcolor{gray}{17.95\%} & \textcolor{gray}{50.29\%} \\ \cmidrule{2-6}
& \multirow{5}{*}{100} & Full FT       & \textcolor{gray}{12.27\%} & \textcolor{gray}{24.94\%} & \textbf{74.57\%} \\
&                      & Prompt Tuning & \textbf{13.05\%} & \textcolor{gray}{33.00\%} & \textcolor{gray}{52.83\%} \\
&                      & Prefix Tuning & \textcolor{gray}{0.68\%} & \textcolor{gray}{45.19\%} & \textcolor{gray}{52.24\%} \\
&                      & LoRA          & \textcolor{gray}{10.96\%} & \textbf{46.15\%} & \textcolor{gray}{74.01\%} \\
&                      & $(IA)^3$      & \textcolor{gray}{6.55\%} & \textcolor{gray}{20.21\%} & \textcolor{gray}{59.53\%} \\ \cmidrule{2-6}
& \multirow{5}{*}{250} & Full FT       & \textbf{26.69\%} & \textcolor{gray}{48.02\%} & \textcolor{gray}{74.61\%} \\
&                      & Prompt Tuning & \textcolor{gray}{22.18\%} & \textcolor{gray}{38.16\%} & \textcolor{gray}{60.06\%} \\
&                      & Prefix Tuning & \textcolor{gray}{14.87\%} & \textcolor{gray}{45.18\%} & \textcolor{gray}{61.13\%} \\
&                      & LoRA          & \textcolor{gray}{20.53\%} & \textbf{55.47\%} & \underline{\textbf{75.33\%}} \\
&                      & $(IA)^3$      & \textcolor{gray}{15.28\%} & \textcolor{gray}{35.25\%} & \textcolor{gray}{65.16\%} \\ \cmidrule{2-6}
& \multirow{5}{*}{500} & Full FT       & \textcolor{gray}{29.33\%} & \textcolor{gray}{58.89\%} & \textcolor{gray}{69.70\%} \\
&                      & Prompt Tuning & \textcolor{gray}{22.09\%} & \textcolor{gray}{40.15\%} & \textcolor{gray}{65.99\%} \\
&                      & Prefix Tuning & \textcolor{gray}{17.72\%} & \textcolor{gray}{46.98\%} & \textcolor{gray}{60.89\%} \\
&                      & LoRA          & \textbf{32.03\%} & \underline{\textbf{59.06\%}} & \textcolor{gray}{69.57\%} \\
&                      & $(IA)^3$      & \textcolor{gray}{9.86\%} & \textcolor{gray}{43.50\%} & \textbf{70.78\%} \\ 
\bottomrule
\end{tabular}
\end{table*}

\subsubsection{Results of RQ3}
\label{subsec:results of RQ3}
According to the results reported in Table \ref{tab:Results of GrahphCodeBERT-Base and DeepSeek-Coder-1.3B-Base with different PEFT methods and full fine-tuning on code smell detection in low-resource scenarios}, it is evident that when the data is limited to 50 samples, the effectiveness of both GraphCodeBERT and DeepSeek-Coder-1.3B with PEFT methods in detecting code smells drops, with \textit{CC} experiencing the most pronounced performance degradation. This significant decline in performance highlights the challenges small and large LMs face in low-resource scenarios. However, when the number of training samples of \textit{CC}, \textit{CM}, or \textit{DC} increases to 500, both models show improved performance compared to when only 50 samples are available. This improvement underscores that a larger dataset enables both small and large LMs using PEFT methods to achieve substantially better performance in code smell detection. Moreover, with 500 samples, GraphCodeBERT and DeepSeekCoder-1.3B with PEFT methods are able to yield reasonably good results compared to training with the original datasets for code smell detection. When the number of training samples reaches 250 or 500, both models with PEFT techniques are able to outperform their fully fine-tuned counterparts.

Additionally, in low-resource scenarios, LoRA tends to be the most effective PEFT method. As indicated by the results of RQ1 (see Section \ref{subsubsec:Best PEFT method}), for GraphCodeBERT, prefix tuning and $(IA)^3$ deliver superior performance, while for DeepSeek-Coder-1.3B, $(IA)^3$ consistently yields the best performance across different code smells. However, when data is scarce, GraphCodeBERT and DeepSeek-Coder-1.3B often attain their highest MCC values through LoRA. It suggests that under different resource settings, the most effective PEFT method may vary.

\noindent\begin{center}
		\begin{tcolorbox}[colback=black!5, colframe=black!20, width=1.0\linewidth, arc=1mm, auto outer arc, boxrule=1.5pt]                       
            {{\textbf{Answer to RQ3:} In low-resource scenarios, the effectiveness of PEFT methods for code smell detection diminishes considerably when data is limited to 50 samples. However, as the number of training samples increases, the performance of PEFT methods shows significant improvement. Moreover, when the number of training samples reaches 250 or 500, several PEFT techniques are able to outperform full fine-tuning, demonstrating their potential even in moderately low-resource settings.
            }}
		\end{tcolorbox}
\end{center}

\subsection{RQ4: What is the effectiveness of PEFT methods compared to state-of-the-art code smell detection approaches?}
\label{subsec:results of RQ4}
\begin{table*}[htbp]
\footnotesize
\centering
\begin{threeparttable}
\caption{Prompt 1 for GPT-4o-mini and DeepSeek-v3 in detecting \textit{CC}, \textit{CM}, \textit{FE}, and \textit{DC}}
\label{tab:prompt template}
\begin{tabular}{m{1.4cm}<{\arraybackslash}m{10.5cm}<{\arraybackslash}}
\toprule
\textbf{Code Smell} & \textbf{Prompt Template*} \\ \cmidrule{1-2}                 
\multirow{5}{*}{CC}    & You are a professional Java programmer. Given a piece of Java code, determine whether it contains ``Complex Conditional'' code smell. If it does, return only ``Yes'', else return ``No''.\\
                    & Definition of ``Complex Conditional'': This code smell occurs when a conditional statement is excessively intricate. \\
                    & \\
                    & \#\#\# Input: [\textit{Code Sample}] \\
                    & \#\#\# Output: [\textit{Yes/No}] \\ \cmidrule{1-2}
\multirow{5}{*}{CM}    & You are a professional Java programmer. Given a piece of Java code, determine whether it contains ``Complex Method'' code smell. If it does, return only ``Yes'', else return ``No''.\\
                    & Definition of ``Complex Method'': This code smell occurs when a method exhibits high cyclomatic complexity. \\
                    & \\
                    & \#\#\# Input: [\textit{Code Sample}] \\
                    & \#\#\# Output: [\textit{Yes/No}] \\ \cmidrule{1-2}
\multirow{5}{*}{FE}    & You are a professional Java programmer. Given a piece of Java code, determine whether it contains ``Feature Envy'' code smell. If it does, return only ``Yes'', else return ``No''.\\
                    & Definition of ``Feature Envy'': This code smell occurs when a method is more interested in a class other than the one it actually is in. \\
                    & \\
                    & \#\#\# Input: [\textit{Code Sample}] \\
                    & \#\#\# Output: [\textit{Yes/No}] \\ \cmidrule{1-2}
\multirow{5}{*}{DC}    & You are a professional Java programmer. Given a piece of Java code, determine whether it contains ``Data Class'' code smell. If it does, return only ``Yes'', else return ``No''.\\
                    & Definition of ``Data Class'': This code smell occurs when a class contains only fields and basic methods for accessing them. \\
                    & \\
                    & \#\#\# Input: [\textit{Code Sample}] \\
                    & \#\#\# Output: [\textit{Yes/No}] \\ 
\bottomrule
\end{tabular}
\begin{tablenotes}
\item * Prompt 1 shown in this table was curated by the authors. Prompt 2 and Prompt 3, which were generated by ChatGPT, are available in our replication package \cite{replicationPackage}.
\end{tablenotes}
\end{threeparttable}
\end{table*}

\begin{figure}
    \centering
    \includegraphics[width=0.9\linewidth]{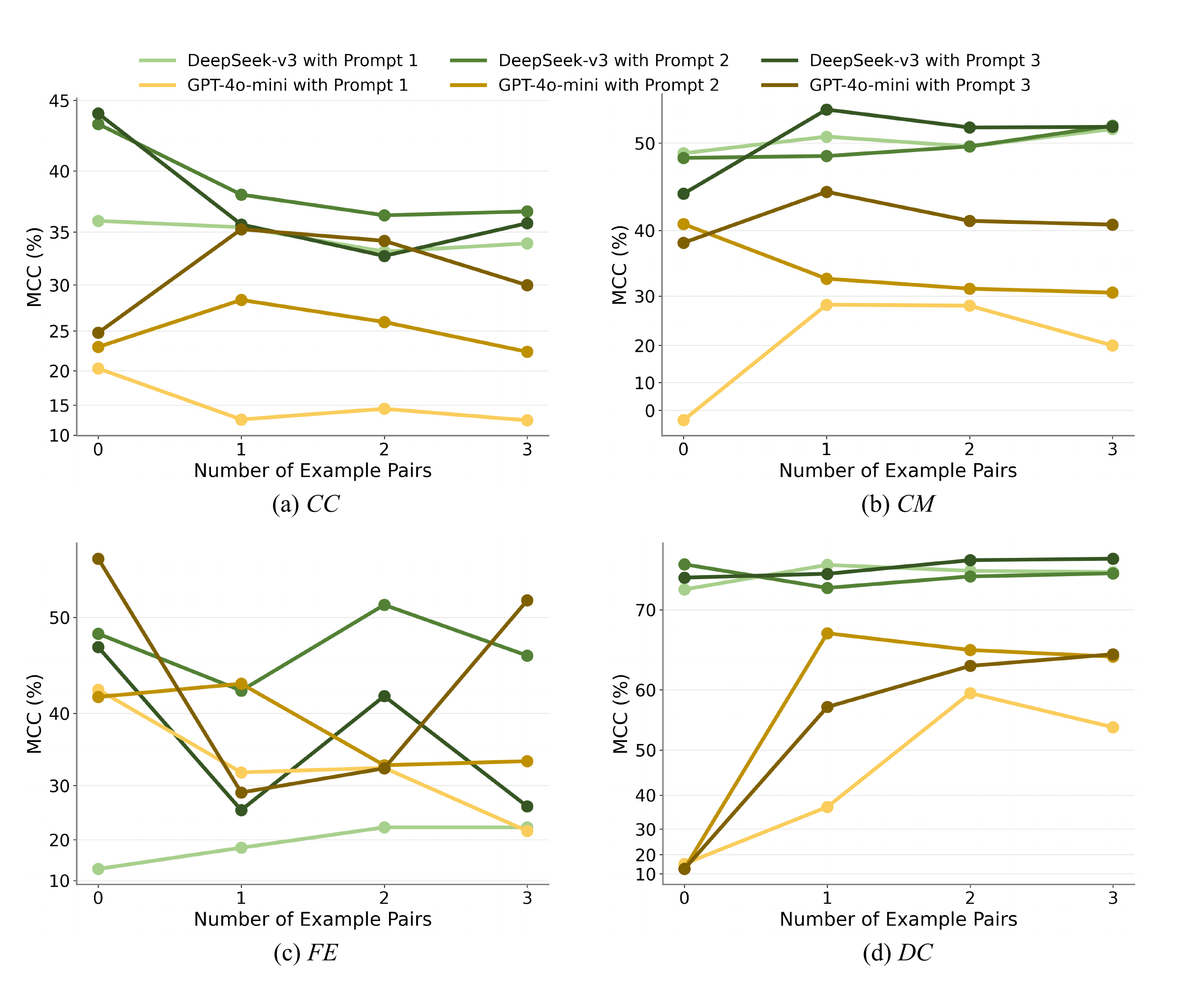}
    \caption{Results of DeepSeek-v3 and GPT-4o-mini on detecting \textit{CC}, \textit{CM}, \textit{FE}, and \textit{DC} under ICL}
    \label{fig:results of ICL}
\end{figure}

\begin{figure}
    \centering
    \includegraphics[width=0.95\linewidth]{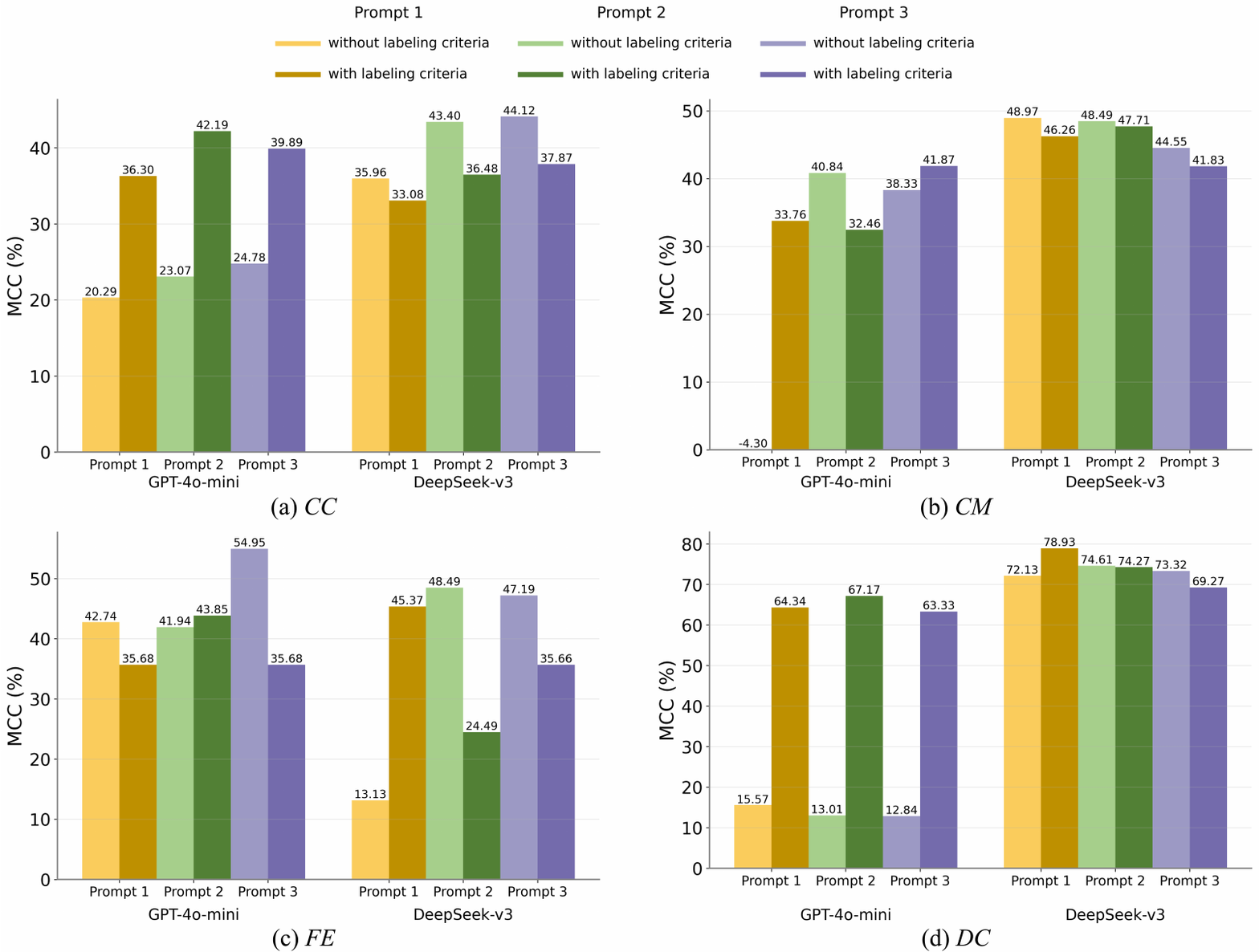}
    \caption{\textcolor{black}{Results of DeepSeek-v3 and GPT-4o-mini on detecting \textit{CC}, \textit{CM}, \textit{FE}, and \textit{DC} with and without labeling criteria}}
    \label{fig:results of labeling criteria}
\end{figure}

\textcolor{black}{To answer this RQ, we benchmark the performance of our PEFT-based methods against the three categories of baselines established in Section \ref{subsec:baselines}: heuristics-based detectors (\textsc{DesigniteJava}, \textsc{PMD}), DL-based approaches (\textsc{DeepSmells}, \textsc{AE-Dense}), and general-purpose LLMs (GPT-4o-mini, DeepSeek-v3) with In-Context Learning (ICL).}

\textcolor{black}{For the general-purpose LLMs, we evaluated their effectiveness under the ICL paradigm. To ensure the robustness of our evaluation against prompt sensitivity, we designed three prompt formulations for each code smell type: a base prompt containing instructional text and definitions (Prompt 1, as shown in Table \ref{tab:prompt template}), and two additional variants (Prompts 2 and 3) generated using ChatGPT \cite{ChatGPT} to introduce linguistic diversity.}

\textcolor{black}{These prompts were applied in three ICL settings:
\begin{enumerate}
    \item \textbf{Zero-shot setting:} prompts used directly without examples;
    \item \textbf{Few-shot setting:} prompts augmented with one, two, or three randomly sampled example pairs (consisting of one smelly and one clean code sample) to provide contextual guidance;
    \item \textbf{Criteria-informed setting:} prompts enriched with the explicit labeling rules derived from our manual review process (Section \ref{subsec: manual review}).
\end{enumerate}}

\textcolor{black}{This multi-faceted design ensures diversity in both prompt formulation and contextual information, thereby supporting a comprehensive evaluation of general-purpose LLMs and enabling a fair comparison with our PEFT-based methods in the context of code smell detection. In total, this yielded 15 unique configurations per smell type (combinations of three prompt variants and five varying contextual levels).}

\textcolor{black}{Figure \ref{fig:results of ICL} visualizes the performance trends of GPT-4o-mini and DeepSeek-v3 across the zero- and few-shot settings, while Figure \ref{fig:results of labeling criteria} illustrates the impact of incorporating labeling criteria. 
Table \ref{tab:Results of RQ4} summarizes the final comparative results, reporting the best-performing configuration of each LLM alongside our PEFT methods and other baselines.}

\subsubsection{\textcolor{black}{Performance of General-Purpose LLMs with ICL}}
\textcolor{black}{It can be observed from Figures \ref{fig:results of ICL} and \ref{fig:results of labeling criteria} that DeepSeek-v3 generally demonstrates better effectiveness in code smell detection compared to GPT-4o-mini across all ICL settings. 
Both models perform better under few-shot settings than in the zero-shot setting; however, their performance does not consistently improve as the number of example pairs increases. Moreover, GPT-4o-mini shows a noticeable performance gain under the criteria-informed setting, whereas DeepSeek-v3 performs better when labeling criteria are not provided. This difference may stem from model characteristics: GPT-4o-mini, a smaller commercial model with stronger instruction alignment, benefits more from explicit and structured guidance such as labeling criteria, which compensates for its limited parameter capacity and contextual reasoning ability. In contrast, DeepSeek-v3, with its larger scale and broader training on code-related corpora, can generalize more effectively from implicit semantics and example patterns, exhibiting stronger reasoning ability even without explicit rules.}

\subsubsection{Overall Comparison with State-of-the-Art Approaches}
\begin{table*}[htbp]
\footnotesize
\centering
\caption{\textcolor{black}{Results of PEFT methods compared to state-of-the-art code smell detection approaches on detecting \textit{CC}, \textit{CM}, \textit{FE}, and \textit{DC} (\textbf{bold}: highest values per code smell)}}
\label{tab:Results of RQ4}
\begin{tabular}{m{1.4cm}<{\arraybackslash}m{4cm}<{\arraybackslash}|m{1.2cm}<{\centering\arraybackslash}|m{1.2cm}<{\centering\arraybackslash}|m{1.2cm}<{\centering\arraybackslash}|m{1.2cm}<{\centering\arraybackslash}}
\toprule
\textbf{Code Smell} & \textbf{Method} & \textbf{Precision} & \textbf{Recall} & \textbf{F1} & \textbf{MCC} \\ \cmidrule{1-6}                    
\multirow{6}{*}{\textit{CC}}
& DesigniteJava                  & 64.01\% & 76.96\% & 69.89\% & 34.41\% \\
& DeepSmells                     & 51.06\% & \textbf{100.00\%} & 67.61\% & 6.63\% \\
& AE-Dense                       & 51.13\% & 49.10\% & 50.10\%  & 2.92\% \\
& GPT-4o-mini                    & \textcolor{black}{72.13\%} & \textcolor{black}{70.11\%} & \textcolor{black}{69.41\%} & \textcolor{black}{42.19\%} \\
& DeepSeek-v3                    & 72.38\% & 71.74\% & 71.53\% & 44.12\% \\
& StarCoderBase-3B with $(IA)^3$ & \textbf{75.18\%} & 74.78\% & \textbf{74.68\%} & \textbf{49.96\%} \\ \cmidrule{1-6}
\multirow{6}{*}{\textit{CM}}
& DesigniteJava                     & 64.69\% & \textbf{98.61\%} & 78.12\% & 52.58\% \\
& DeepSmells                        & 63.64\% & 79.87\% & 70.83\% & 33.35\% \\
& AE-Dense                          & 50.54\% & 52.05\% & 51.29\% & 3.05\% \\
& GPT-4o-mini                       & 79.06\% & 67.25\% & 63.54\% & 44.78\% \\
& DeepSeek-v3                       & 79.74\% & 73.78\% & 72.40\% & 53.19\% \\
& DeepSeek-Coder-6.7B with $(IA)^3$ & \textbf{83.52\%} & 83.36\% & \textbf{83.34\%} & \textbf{66.88\%} \\ \cmidrule{1-6}
\multirow{6}{*}{\textit{FE}}
& DesigniteJava                     & 73.91\% & 54.84\% & 62.96\% & 36.73\% \\
& DeepSmells                        & \textbf{97.10\%} & 51.54\% & 67.34\% & 7.48\% \\
& AE-Dense                          & 50.82\% & \textbf{100.00\%} & 67.39\% & 12.80\% \\
& GPT-4o-mini                       & 77.53\% & 77.42\% & 77.40\% & 54.95\% \\
& DeepSeek-v3                       & \textcolor{black}{77.00\%} & \textcolor{black}{74.19\%} & \textcolor{black}{73.50\%} & \textcolor{black}{51.12\%} \\
& DeepSeek-Coder-1.3B with $(IA)^3$ & 83.87\% & 83.87\% & \textbf{83.87\%} & \textbf{67.74\%} \\ \cmidrule{1-6}
\multirow{6}{*}{\textit{DC}}
& PMD                              & 76.29\% & 89.74\% & 82.47\% & 62.83\% \\
& DeepSmells                       & 87.72\% & 67.68\% & 76.41\% & 49.55\% \\
& AE-Dense                         & 52.79\% & \textbf{94.36\%} & 67.70\% & 16.18\% \\
& GPT-4o-mini                      & 84.87\% & 82.51\% & 82.21\% & 67.35\% \\
& DeepSeek-v3                      & \textcolor{black}{89.48\%} & \textcolor{black}{89.45\%} & \textcolor{black}{89.45\%} & \textcolor{black}{78.93\%} \\
& GraphCodeBERT with prefix tuning & \textbf{90.95\%} & 90.03\% & \textbf{89.97\%} & \textbf{79.26\%} \\ 
\bottomrule
\end{tabular}
\end{table*}
Based on the results presented in Table \ref{tab:Results of RQ4}, we can find that LMs fine-tuned with PEFT techniques consistently demonstrate the highest effectiveness among all existing methods for code smell detection. \textcolor{black}{Specifically, compared to the best-performing baseline, PEFT-tuned LMs achieve MCC improvements of 5.85\%, 13.69\%, 12.79\%, and 0.33\% for \textit{CC}, \textit{CM}, \textit{FE}, and \textit{DC}, respectively. Among various baseline approaches, DeepSeek-v3 achieves the most competitive overall performance. However, despite its relative strength in detecting \textit{CC} and \textit{DC}, a significant performance gap remains: DeepSeek-v3 lags considerably behind PEFT-tuned LMs in \textit{CM} and \textit{FE} detection, where the performance differences exceed 10\%.}

Furthermore, our results indicate that traditional DL-based methods (\textsc{DeepSmells} and \textsc{AE-Dense}) exhibit substantially inferior performance compared to both PEFT and the other baseline approaches.
This finding underscores the limitations of conventional DL models in code smell detection, particularly when compared to more advanced techniques like PEFT and ICL.

\noindent\begin{center}
		\begin{tcolorbox}[colback=black!5, colframe=black!20, width=1.0\linewidth, arc=1mm, auto outer arc, boxrule=1.5pt]                       
            {{\textbf{Answer to RQ4:} In comparison with existing code smell detection approaches, LMs tuned with PEFT methods reveal notable performance improvements, highlighting the effectiveness of PEFT methods for code smell detection.}}
		\end{tcolorbox}
\end{center}

\section{Discussion}
\label{sec:discussion}
\noindent \textbf{Approaches with improved effectiveness for \textit{CC} code smell detection are needed:} According to Fowler \cite{fowler1999refactoring}, four factors contribute to making programs difficult to work with, one of which is \textit{Complex Conditional} (\textit{CC}) logic that complicates modification. The presence of \textit{CC} in source code hinders testing and maintenance \cite{fowler1999refactoring, liu2014automated, alhefdhi2022framework}. Thus, identifying \textit{CC} code smells in software can significantly improve code quality \cite{liu2014automated}. However, the effectiveness of existing methods for detecting this type of code smell is considerably limited. As shown in Table \ref{tab:Results of RQ4}, existing methods, including heuristics-based (e.g., \textsc{DesigniteJava}), DL-based (e.g., \textsc{DeepSmells}) and \textsc{AE-Dense}), and general-purpose LLMs under ICL settings (e.g., GPT-4o-mini and DeepSeek-v3), all demonstrate limited success in detecting \textit{CC}, with all MCC values below 50\%, and some even as low as 6.63\%. Although our proposed method, StarCoderBase-3B with $(IA)^3$, improved the performance of \textit{CC} code smell detection significantly, reaching an F1-score of 74.68\% and a MCC value of 49.96\%, it still falls short of expectations due to challenges such as the considerable existence of false positives and negatives, which may limit its practical applicability. Accurate identification of \textit{CC} code smells can help developers better refactor the code, ultimately improving the maintainability of the software system. Despite the progress made, the detection of \textit{CC} remains a challenging task that requires further improvements.

\vspace{0.1cm}
\noindent \textbf{PEFT methods can enhance the capacity of LMs in class-level code understanding:} In this study, we selected four representative types of code smells, i.e., \textit{CC}, \textit{CM}, \textit{FE}, and \textit{DC}, to evaluate the effectiveness of PEFT methods. While only \textit{DC} is formally categorized as a class-level smell, \textit{FE} detection also inherently requires class-level analysis as well, as it involves analyzing method-class dependencies (see Section \ref{subsubsec:selected code smells} and Section \ref{subsubsec:Data preparation}). According to the recent studies \cite{silva2024detecting, mesbah2025leveraging}, LLMs not specifically fine-tuned for code smell detection generally perform worse when tasked with identifying smells that demand class-level understanding.
This aligns with our intuitive expectation that class-level code understanding is more difficult, as it requires capturing higher-level structural relationships (e.g., field-method interactions, class cohesion, and design-level abstractions).
Despite this challenge, most existing research on PEFT methods has predominantly concentrated on the their perfomance for method-level code understanding \cite{liu2023empirical, wang2023one, liu2024delving}, leaving their effectiveness on class-level tasks largely unexplored.
Our work bridges this critical gap by evaluating PEFT methods on code smell detection, a task spans both method- and class-level analysis. Surprisingly, as discussed in Section \ref{subsubsec:code understanding}, our results reveal that LLMs with PEFT methods not only overcome the anticipated class-level challenges but even outperform method-level detection in some cases. These findings suggest that PEFT methods not only improve parameter efficiency but also strengthen the proficiency of LLMs to perform robust code understanding across different levels of granularity.

\vspace{0.1cm}
\noindent \textcolor{black}{\textbf{When fine-tuning LMs for code smell detection, more trainable parameters do not guarantee better results:} Our findings indicate that the effectiveness of fine-tuning is not primarily determined by the number of parameters being updated. Although full fine-tuning adjusts all model parameters, it does not consistently outperform PEFT methods in code smell detection. For many SLMs and most code smell types, PEFT approaches achieve comparable or even superior performance while adding only a small fraction of parameters. This indicates that large-scale parameter updates are not necessary for effective model adaptation in this task.
A similar pattern emerges when comparing individual PEFT techniques. LoRA and prefix tuning introduce far more trainable parameters than prompt tuning or $(IA)^3$, yet their additional capacity does not reliably translate into better performance. In some cases, LoRA even performs markedly worse despite being the most parameter-intensive PEFT method. In contrast, $(IA)^3$ frequently provides the best overall performance while relying on relatively few added parameters. 
Taken together, these observations confirm that increasing the number of trainable parameters alone does not guarantee better outcomes. Instead, the effectiveness of fine-tuning depends on how efficiently the model’s internal representations are adapted and on the strategic placement of parameter updates for code smell detection.}

\vspace{0.1cm}
\noindent \textbf{While larger LMs with PEFT methods often achieve better effectiveness in code smell detection, SLMs still retain distinct advantages:} 
The results of RQ1 reveal that LLMs generally outperform SLMs in code smell detection, particularly in identifying \textit{FE}. This performance advantage aligns with established neural scaling laws \cite{kaplan2020scaling}, which posit that model capabilities improve predictably with increased scale. 
However, SLMs are able to achieve comparable performance in detecting \textit{CC} and \textit{CM}, and even significantly outperform LLMs in the detection of \textit{DC}. Moreover, as illustrated in Figure \ref{fig:peak gpu}, SLMs exhibit substantially lower peak GPU memory consumption compared to LLMs, making them a more resource-efficient choice. These findings highlight a critical trade-off: while LLMs tend to yield higher absolute performance, SLMs offer a competitive and cost-effective alternative, especially in settings with limited computational resources.

\vspace{0.1cm}
\noindent \textbf{Considering the model used, the code smell type, and the availability of resources when selecting the appropriate PEFT method:} Based on the results of RQ1 and RQ3, we can find that for code smell detection, the most effective PEFT method can differ significantly depending on the models and the amount of available data. From Table \ref{tab:results_of_RQ1_CC_CM} and Table \ref{tab:results_of_RQ1_FE_DC}, we observe that although $(IA)^3$ generally achieves the best performance for LLMs across most code smell types, this is not universally the case. For certain models, particularly SLMs, and specific types of code smells, prefix tuning or even prompt tuning may outperform $(IA)^3$. As indicated in Figure \ref{fig:peak gpu}, the relatively high peak GPU memory consumption of $(IA)^3$ may limit its practicality in constrained environments, where more lightweight PEFT methods could be more suitable. Moreover, in low-resource scenarios where training data is limited, LoRA emerges as the most effective choice (see results of RQ3 in Section \ref{subsec:results of RQ3}). These findings emphasize the importance of selecting PEFT strategies holistically, taking into account not only the model size and available resources, including both GPU memory and training data, but also the specific demands of the code smell being detected.

\vspace{0.1cm}
\noindent \textbf{Start model selection for code smell detection with a smaller dataset to optimize resource usage:}
In low-resource scenarios, even when the dataset size was drastically reduced, the models' effectiveness did not degrade significantly, and they still yielded promising results, as shown in the results of RQ3. This finding indicates that, even with a significantly smaller dataset, both small and large LMs can maintain robust performance for the code smell detection task. Consequently, beginning fine-tuning experiments with a smaller-size dataset can be an effective strategy to optimize resource usage. This approach allows researchers and practitioners to initially evaluate a wide range of LMs in a shorter time and with fewer computational resources, saving substantial effort. It enables them to identify the most suitable model for the specific task through an initial evaluation before committing to the more resource-intensive process of training on larger datasets. Additionally, it helps in assessing the trade-offs between resource consumption and performance, providing insight into the models' scalability. By leveraging smaller datasets for initial evaluation, one can make informed decisions about which models are likely to perform well with limited resources, and further fine-tune or scale up as necessary, without wasting valuable time or computational power.

\vspace{0.1cm}
\noindent \textbf{ICL shows promise as an effective alternative to PEFT methods for certain types of code smells:} \textcolor{black}{In RQ4, we investigated the performance of general-purpose LLMs in code smell detection under In-Context Learning (ICL) settings, including zero-shot, few-shot, and criteria-informed setups. As shown in Table \ref{tab:Results of RQ4}, for \textit{CC} and \textit{DC}, the optimal ICL configuration of DeepSeek-v3 achieves MCC scores only 1\%–5\% lower than those attained by the top-performing PEFT-tuned LMs. This suggests that, for certain types of code smells, well-crafted prompts with ICL may enable general-purpose LLMs to produce competitive results. Moreover, applying ICL technique requires no model training or parameter updates \cite{zhou2024large, tom2020language, li2025large}, which is particularly appealing for resource-constrained scenarios or when model weights cannot be modified (e.g., in commercial LLM APIs). However, for \textit{CM} and \textit{FE}, ICL lags significantly behind, with performance deficits of up to 13.69\% and 16.62\% in MCC, respectively, compared to our PEFT-based approaches. These results indicate that detecting these smell types likely requires deeper model adaptation to capture subtle semantic patterns and structural nuances that few-shot prompting alone cannot convey. Overall, while ICL offers a lightweight and flexible alternative, especially for rapid prototyping or scenarios with limited computational resources, PEFT techniques remain the more robust and reliable solution for achieving high-accuracy code smell detection across a broad range of smell types.}

\vspace{0.1cm}
\noindent \textbf{Future work could explore the use of Retrieval-Augmented Generation (RAG) for code smell detection:} According to Figure \ref{fig:results of ICL}, while models generally perform better in code smell detection when provided with randomly selected examples compared to the zero-shot setting, their performance does not consistently improve as the number of example pairs increases. This observation suggests that simply increasing the number of in-context examples without considering their quality, relevance, or diversity may not be sufficient to enhance model effectiveness for the task of code smell detection. By contrast, Retrieval-Augmented Generation (RAG) \cite{lewis2020retrieval} can dynamically retrieve the most relevant examples from an external knowledge base to augment the prompts. By grounding predictions in high-quality and contextually appropriate evidence, RAG has the potential to enhance detection accuracy while avoiding the need for fine-tuning or complex prompt engineering. As such, future research could explore whether RAG-based approaches can provide a more effective and scalable solution to PEFT methods for code smell detection.

\section{Threats to Validity}
\label{sec:threats}

\textbf{Construct Validity:} One primary threat to the construct validity of this study is the construction of the datasets. We first used \textsc{DesigniteJava} \cite{designitejava} to detect potential positive and negative samples, then manually reviewed all the candidate data to validate the datasets, ensuring there were no false positives or negatives.

First, relying solely on \textsc{DesigniteJava} and \textsc{PMD} for initial sample detection may introduce bias. However, \textsc{DesigniteJava} and \textsc{PMD} are well-established and widely adopted tools in Java code smell detection \cite{chaniotaki2021architecture, sharma2021qscored, shah2023mining, tushar2016does, taibi2017how}, providing a robust foundation for the identification of potential code smells. We also acknowledge that future studies could benefit from incorporating additional tools in the initial code smell detection process, which may help mitigate the threat to construct validity by providing more diverse perspectives on the data.

Second, the manual validation process, while aimed at ensuring dataset accuracy, could introduce some subjectivity due to its reliance on human judgment. To minimize this threat, we conducted separate pilot validations for the four types of code smells before formally annotating all the data samples. During this phase, we established and refined clear identification criteria to ensure alignment between annotators. This approach not only improved labeling consistency but also reduced potential bias in the dataset.

Lastly, due to budget constraints for commercial API usage, our evaluation of general-purpose LLMs (GPT-4o-mini and DeepSeek-v3) under ICL settings in RQ4 was limited to a maximum of three example pairs. This sample size may not fully represent the upper bounds of ICL performance. Future work could expand on this by increasing the number of example pairs and evaluating additional models to better understand the full potential of ICL performance. However, prior studies such as UniLog \cite{xu2024unilog} have shown that the performance of LLMs on software engineering tasks does not always improve linearly with more examples, which suggests that while increasing the number of examples may be beneficial, performance gains are not guaranteed. As such, we believe our current evaluation still provides valuable insights and that this threat to validity is partially mitigated.

\textbf{External Validity:} In this study, we evaluated the effectiveness of state-of-the-art PEFT methods using 4 small LMs and 5 LLMs. However, due to limited computational resources, we did not include larger models with over 7B parameters, which may impact the generalizability of our findings. 

\textcolor{black}{Furthermore, our study focused exclusively on four widely used and representative PEFT methods (i.e., prompt tuning, prefix tuning, LoRA, and $(IA)^3$) implemented within the Hugging Face PEFT library. Techniques such as adapter tuning \cite{houlsby2019parameter} were excluded because they are not yet well supported by the library and would require additional external implementations, which could compromise the consistency of our experimental setup. As a result, our findings may not fully generalize to PEFT methods beyond the four examined ones. To mitigate this limitation, we carefully selected diverse PEFT methods covering the major families (prompt-based, LoRA-based, and $(IA)^3$) to ensure representativeness across categories. In future work, we plan to incorporate adapter tuning and other emerging PEFT techniques as they become more maturely integrated into the Hugging Face ecosystem, thereby enabling a broader and more comprehensive comparison.}

Moreover, our study applied PEFT methods exclusively to the detection of Java code smells, focusing on four specific types: three method-level smells (\textit{Complex Conditional}, \textit{Complex Method}, and \textit{Feature Envy}) and one class-level smell (\textit{Data Class}). While this selection provides a certain degree of diversity, it does not fully capture the wide spectrum of code smell types encountered in real-world software systems. Expanding future work to include a wider range of smell categories would enhance the comprehensiveness of the evaluation. In addition, our current work is limited to Java. Applying PEFT techniques to detect code smells in other programming languages, such as C\# or Python, could yield further insights into their generalizability and adaptability across different language ecosystems.


\textbf{Internal Validity:} The settings of hyper-parameters for the four PEFT methods and full fine-tuning play a crucial role in the performance of the models. Variations in these settings could lead to different results, potentially affecting the reliability of our findings. To mitigate this threat, we referred to previous works and conducted pilot experiments to determine the appropriate hyper-parameter values. 

\textcolor{black}{In our experiments, we used a batch size of 1 for all models due to computational resource constraints, which ensured consistency but could potentially affect convergence and performance. To evaluate this, we conducted supplementary experiments with different batch sizes, reported in Appendix \ref{sec:batch size}, which show that batch size has a limited impact on overall detection performance across PEFT methods, suggesting the main conclusions remain robust. Similarly, each configuration was run only once using a fixed random seed (42), which does not fully capture the stochasticity inherent in training. Additional runs with two alternative seeds (0 and 999) for a representative case (GraphCodeBERT on \textit{CC} detection), reported in Appendix \ref{sec:seed}, indicate minor performance variations, confirming robustness to random initialization. Nevertheless, we acknowledge that more comprehensive experiments with larger batch sizes and multiple seeds across more models could provide further insights and increase statistical reliability.}

\textcolor{black}{CodeBERT, GraphCodeBERT, and CodeT5 have a maximum input length of 512 tokens, requiring longer code samples to be truncated. This could potentially omit semantically important content relevant for code smell detection. We analyzed token length distributions and manually reviewed a subset of long samples (50 positives and 50 negatives per smell type) in Appendix \ref{sec:impact of token length}. The review showed that essential indicators of each code smell typically occur within the first 512 tokens, and truncated segments rarely contain critical information. Thus, while context window limitations represent a potential threat, their impact on overall evaluation results is likely minimal.}

Additionally, data leakage represents another potential threat to internal validity. First, the poor zero-shot performance of GPT-4o-mini in detecting \textit{CC} and \textit{CM} code smells suggests that the model was not exposed to the test data (see Section \ref{subsec:results of RQ4}), as we would expect stronger zero-shot effectiveness if GPT-4o-mini had memorized the data. Second, the significant improvements in performance for both small and large LMs following the application of PEFT techniques further suggest that memorization is unlikely to be an issue. Such substantial performance gains would not occur if the models had already memorized the test sets. 
Furthermore, Cao \textit{et al.} \cite{cao2024concerned} highlight that code LMs do not consistently exhibit reduced performance on data extending beyond their training cut-off dates. 
This careful management of data, along with the observed performance improvements after fine-tuning, ensures that data leakage was not a major threat, thereby maintaining the internal validity of our findings. 

\section{Related Work}
\label{sec:related work}
\subsection{Code Smell Detection}
Existing approaches to code smell detection predominantly rely on heuristics-based methods, which utilize predefined rules to identify problematic code patterns \cite{pmd, designitejava, campbell2013sonarqube, fokaefs2007jdeodorant, vidal2015jspirit, moha2010decor}. These methods aim to flag potential smells by analyzing the source code against established standards. For instance, PMD \cite{pmd}, a widely used static code analysis tool, supports 16 programming languages and applies rule sets to detect code smells. These rule sets can be tailored to suit specific project requirements or design standards, making the tool adaptable for different contexts. Similarly, Sharma \cite{designitejava} introduced \textsc{DesigniteJava}, an advanced tool capable of identifying code smells at multiple levels — ranging from architecture and design to implementation. 

Traditional heuristics-based code smell detectors face inherent challenges, such as difficulty in defining effective thresholds for detection. In response, Machine Learning (ML) and Deep Learning (DL) techniques have gained traction as promising alternatives. Maiga \textit{et al.} \cite{maiga2012support} implemented SVMDetect based on an ML technique — support vector machines. Fontana \textit{et al.} \cite{fontana2016comparing} evaluated 16 ML algorithms across four types of code smells using 74 software systems, reporting that J48 and Random Forest delivered the highest performance.

Liu \textit{et al.} \cite{liu2021deep} utilized deep neural networks and advanced DL techniques to perform useful feature selection from source code and map these features to labels (smelly or not). Sharma \textit{et al.} \cite{sharma2021code} trained Convolution Neural Network (CNN), Recurrent Neural Network (RNN), and autoencoder DL models in different configurations to detect code smells. Their work confirms the feasibility of applying transfer-learning for code smell detection. Ho \textit{et al.} \cite{ho2023fusion} developed DeepSmells based on a CNN and long short-term memory (LSTM) networks for detecting code smells. They conducted an empirical study to compare DeepSmells with other well-established detection tools. Their results demonstrate that DeepSmells outperforms the baselines, achieving superior performance. They also anticipated that leveraging CodeBERT would enhance the effectiveness of code smell detection. 

We conduct an extensive analysis of the Machine Learning Code Quality (MLCQ) dataset, focusing on how these LLMs perform when prompted to identify and classify code smells. 

Recently, researchers have started to explore the use of LLMs for code smell detection. Mesbah \textit{et al.} \cite{mesbah2025leveraging} used the MLCQ dataset to analyze how GPT-4 and LLaMA perform when prompted to identify and classify code smells. Silva \textit{et al.} \cite{silva2024detecting} employed two types of prompts to query ChatGPT for identifying \textit{Blob}, \textit{Data Class}, \textit{Feature Envy}, and \textit{Long Method} code smells. Sadik \textit{et al.} \cite{sadik2025benchmarking} created a dataset for evaluating LLMs on automated code smell detection. They benchmarked two state-of-the-art LLMs and found that LLMs are not yet ready to replace deterministic tools in all aspects of code quality assurance. Wu \textit{et al.} \cite{wu2024ismell} trained a Mixture of Experts (MoE) model, leveraging specialized expert toolsets for code smell detection. They subsequently utilized the detection results to prompt LLMs for refactoring.

Most prevailing ML-based or DL-based approaches for code smell detection train models from scratch. On the other hand, existing LLM-based approaches typically rely on simple prompting strategies without any form of fine-tuning or adaptation.
Compared to their works, we harness the power of LLMs using PEFT methods to detect four types of widely studied code smells (\textit{Complex Conditional}, \textit{Complex Method}, \textit{Feature Envy}, and \textit{Data Class}), aiming to enhance the models' ability to capture the nuanced relationships between different code smell types.

\subsection{PEFT Methods on Software Engineering Tasks}
Several studies have evaluated PEFT methods on various Software Engineering (SE) tasks. Wang \textit{et al.} \cite{wang2023one} fine-tuned UniXcoder and CodeT5 by inserting the parameter-efficient structure adapter for code search and summarization. According to their results, adapter tuning has shown effectiveness in cross-lingual and low-resource scenarios, surpassing full fine-tuning and effectively mitigating catastrophic forgetting. Liu \textit{et al.} \cite{liu2023empirical} carried out experiments by fine-tuning code models with adapter, prefix tuning, LoRA, and MHM \cite{he2022towards} on both the code understanding task and code generation task. Their findings indicate that adapter tuning and LoRA can achieve promising performance on these two SE tasks with reduced training costs and memory requirements. Liu \textit{et al.} \cite{liu2024delving} examined whether adapter tuning and LoRA can perform well on code-change-related tasks, specially just-in-time defect prediction and commit message generation. Their study demonstrates that PEFT methods can serve as a powerful alternative to full fine-tuning for dynamic code comprehension. Weyssow \textit{et al.} \cite{weyssow2023exploring} compared PEFT methods, including prompt tuning and others, with In-Context Learning (ICL) in terms of code generation. Their investigation reveals the superiority of PEFT over ICL and highlights the potential opportunities of tuning LLMs in diverse SE scenarios. Li \textit{et al.} \cite{li2024comprehensive} created an instruction dataset of Automated Program Repair (APR) using prompt engineering. They employed this dataset to fine-tune four LLMs using four different PEFT methods. The results show that $(IA)^3$ achieves the highest fixing ability compared to prompt tuning, prefix tuning, and LoRA.

These studies focus on evaluating PEFT methods across a range of SE tasks, primarily at the method level. Compared to their works, our work specifically targets the task of code smell detection, which, to the best of our knowledge, has not been studied before. Additionally, our work further explores the capacity of LLMs on class-level code understanding, thereby addressing a significant gap in existing research.

\section{Conclusions and Future Work}
\label{sec:conclusions}
In this work, we investigated the effectiveness of various PEFT methods on different LMs for code smell detection. We also evaluated how the performance of these PEFT methods varies under different conditions, such as varying hyper-parameter settings and the availability of training data. Our analysis reveals that PEFT methods are more efficient than full fine-tuning and In-Context Learning (ICL) in most cases, underscoring the potential of PEFT methods to enhance code smell detection and offer valuable insights for optimizing LMs for other Software Engineering (SE) tasks. Additionally, we have constructed high-quality datasets for four widely studied code smells, i.e., \textit{Complex Conditional}, \textit{Complex Method}, \textit{Feature Envy}, and \textit{Data Class}, and we provided the replication package of this work online \cite{replicationPackage} to facilitate future research and ensure the reproducibility of our study results.

In the next step, we plan to further explore how PEFT performs in cross-language scenarios for code smell detection. Moreover, we plan to investigate the application of PEFT methods for fine-tuning LMs to assist in a wider range of SE tasks, such as bug detection, code completion, and code smell refactoring. These directions will help us better understand the generalizability and practical value of PEFT methods in real-world software development settings.


\section*{Data Availability}
The replication package for this work has been made available at \url{https://github.com/MabelQi/PEFT4CSD}.

\begin{acks}
This work has been partially supported by the National Natural Science Foundation of China (NSFC) with Grant No. 92582203 and the Major Science and Technology Project of Hubei Province under Grant No. 2024BAA008. The numerical calculations in this paper have been done on the supercomputing system in the Supercomputing Center of Wuhan University.
\end{acks}

\bibliographystyle{ieeetr}
\bibliography{references}

\section*{\textcolor{black}{Appendix}}
\appendix

\section{\textcolor{black}{The Impact of Other Hyper-Parameters on PEFT Methods}}
\textcolor{black}{In addition to the primary experimental settings discussed in the main text, we investigate the effects of two additional hyper-parameters on PEFT methods: batch size and random seed selection. Understanding the sensitivity of PEFT methods to these factors is crucial for assessing the stability and reproducibility of our results.}

\textcolor{black}{Specifically, we examine how varying batch sizes during training influences model efficiency and prediction accuracy. We also evaluate performance variations across multiple random seeds to ensure that our findings are not driven by a particular initialization.}

\subsection{\textcolor{black}{Batch Size}}
\label{sec:batch size}
\begin{table}[htbp]
\footnotesize
\textcolor{black}{
\centering
\caption{\textcolor{black}{Results of GraphCodeBERT with PEFT methods and full fine-tuning on detecting \textit{CC} across different batch sizes}}
\label{tab:Results of PEFT with different batch sizes}
\begin{tabular}{m{1.5cm}<{\centering}m{1.8cm}<{\arraybackslash}|m{1.2cm}<{\centering\arraybackslash}|m{1.2cm}<{\centering\arraybackslash}|m{1.2cm}<{\centering\arraybackslash}|m{1.2cm}<{\centering\arraybackslash}}
\toprule
\textbf{Batch Size} & \textbf{Method} & \textbf{Precision} & \textbf{Recall} & \textbf{F1} & \textbf{MCC} \\ \cmidrule{1-6}                    
\multirow{5}{*}{1}
& Full FT       & 74.06\% & 74.02\% & 74.01\% & 48.08\% \\
& Prompt Tuning & 65.49\% & 65.33\% & 65.24\% & 30.81\% \\
& Prefix Tuning & 70.08\% & 69.89\% & 69.82\% & 39.97\% \\ 
& LoRA          & 73.15\% & 73.15\% & 73.15\% & 46.30\% \\
& $(IA)^3$      & 72.50\% & 72.50\% & 72.50\% & 45.00\% \\ \cmidrule{1-6}
\multirow{5}{*}{8}
& Full FT       & 72.10\%  & 72.07\%  & 72.06\% & 44.16\% \\
& Prompt Tuning & 63.20\%  & 63.15\%  & 63.12\% & 26.36\% \\
& Prefix Tuning & 71.24\%  & 70.98\%  & 70.89\% & 42.22\% \\ 
& LoRA          & 73.76\%  & 73.59\%  & 73.54\% & 47.34\% \\
& $(IA)^3$      & 72.01\%  & 71.85\%  & 71.80\% & 43.85\% \\ \cmidrule{1-6}
\multirow{5}{*}{16}
& Full FT       & 73.19\% & 72.83\% & 72.72\% & 46.02\%  \\
& Prompt Tuning & 63.48\% & 63.48\% & 63.47\% & 26.96\%  \\
& Prefix Tuning & 69.66\% & 69.02\% & 68.77\% & 38.68\%  \\ 
& LoRA          & 73.59\% & 73.59\% & 73.59\% & 47.18\%  \\
& $(IA)^3$      & 71.65\% & 71.30\% & 71.19\% & 42.95\%  \\ \cmidrule{1-6}
\multirow{5}{*}{32}
& Full FT       & 73.08\% & 72.61\% & 72.47\% & 45.69\%  \\
& Prompt Tuning & 63.27\% & 63.26\% & 63.25\% & 26.53\%  \\
& Prefix Tuning & 69.32\% & 68.48\% & 68.13\% & 37.79\%  \\ 
& LoRA          & 73.46\% & 73.15\% & 73.06\% & 46.61\%  \\
& $(IA)^3$      & 70.66\% & 69.35\% & 68.85\% & 39.99\%  \\
\bottomrule
\end{tabular}
}
\end{table}
\textcolor{black}{Table \ref{tab:Results of PEFT with different batch sizes} presents the performance of GraphCodeBERT with different PEFT methods (prompt tuning, prefix tuning, LoRA, and $(IA)^3$), as well as full fine-tuning, for detecting the \textit{Complex Conditional} (\textit{CC}) code smell across various batch sizes (1, 8, 16, and 32).}

\textcolor{black}{From Table \ref{tab:Results of PEFT with different batch sizes}, we observe that batch size has a limited impact on overall detection performance. Full fine-tuning and LoRA consistently achieve the highest MCC values across all batch sizes, while $(IA)^3$ also maintains strong performance. Prompt tuning generally yields lower scores, and prefix tuning falls in between. Minor fluctuations are present; for example, LoRA attains slightly higher F1 and MCC scores at batch size 8 compared to batch size 1. Overall, the performance trends remain stable, suggesting that PEFT methods are relatively robust to batch size variations and that the main experimental conclusions are unlikely to be materially affected.
}

\begin{table}[htbp]
\footnotesize
\centering
\textcolor{black}{
\caption{\textcolor{black}{Results of GraphCodeBERT with PEFT methods and full fine-tuning on detecting \textit{CC} across different seeds}}
\label{tab:Results of PEFT with different seeds}
\begin{tabular}{m{1.5cm}<{\centering}m{1.8cm}<{\arraybackslash}|m{1.2cm}<{\centering\arraybackslash}|m{1.2cm}<{\centering\arraybackslash}|m{1.2cm}<{\centering\arraybackslash}|m{1.2cm}<{\centering\arraybackslash}}
\toprule
\textbf{Seed} & \textbf{Method} & \textbf{Precision} & \textbf{Recall} & \textbf{F1} & \textbf{MCC} \\ \cmidrule{1-6}                    
\multirow{5}{*}{0}
& Full FT       & 73.47\%  & 72.17\%  & 71.79\% & 45.62\% \\
& Prompt Tuning & 64.89\%  & 64.89\%  & 64.89\% & 29.78\% \\
& Prefix Tuning & 71.32\%  & 71.30\%  & 71.30\% & 42.62\% \\ 
& LoRA          & 73.17\%  & 73.04\%  & 73.01\% & 46.21\% \\
& $(IA)^3$      & 72.44\%  & 72.17\%  & 72.09\% & 44.61\% \\ \cmidrule{1-6}
\multirow{5}{*}{42}
& Full FT       & 74.06\% & 74.02\% & 74.01\% & 48.08\% \\
& Prompt Tuning & 65.49\% & 65.33\% & 65.24\% & 30.81\% \\
& Prefix Tuning & 70.08\% & 69.89\% & 69.82\% & 39.97\% \\ 
& LoRA          & 73.15\% & 73.15\% & 73.15\% & 46.30\% \\
& $(IA)^3$      & 72.50\% & 72.50\% & 72.50\% & 45.00\% \\ \cmidrule{1-6}
\multirow{5}{*}{999}
& Full FT       & 73.19\% & 72.83\% & 72.72\% & 46.02\%  \\
& Prompt Tuning & 64.26\% & 64.24\% & 64.22\% & 28.50\%  \\
& Prefix Tuning & 71.21\% & 69.57\% & 68.96\% & 40.74\%  \\ 
& LoRA          & 73.07\% & 72.93\% & 72.90\% & 46.00\%  \\
& $(IA)^3$      & 72.23\% & 72.17\% & 72.15\% & 44.41\%  \\
\bottomrule
\end{tabular}
}
\end{table}
\subsection{\textcolor{black}{Random Seed}}
\label{sec:seed}
\textcolor{black}{Table \ref{tab:Results of PEFT with different seeds} presents the results of GraphCodeBERT with different PEFT methods (i.e., prompt tuning, prefix tuning, LoRA, and $(IA)^3$), as well as full fine-tuning, for detecting the \textit{Complex Conditional} (\textit{CC}) code smell across three different random seeds (0, 42, and 999).}

\textcolor{black}{As shown in Table \ref{tab:Results of PEFT with different seeds}, the variations in performance metrics across different random seeds are minor. These small differences indicate that our core findings are robust to the effects of random initialization, with overall trends remaining consistent. While some fluctuations in performance are observed, they do not materially alter the conclusions drawn from the experiments. Nevertheless, we acknowledge that conducting multiple runs with different seeds would yield more statistically reliable results.}

\section{\textcolor{black}{The Impact of Context Window Limit to SLMs}}
\label{sec:impact of token length}
\textcolor{black}{In the process of data construction, considering our GPU memory limitations, we filtered out any samples that exceed 1,024 tokens. However, several SLMs used in our study, including CodeBERT, GraphCodeBERT, and CodeT5, can only accept input sequences up to 512 tokens (see Section \ref{sec:base model selection}). Consequently, any code sample longer than 512 tokens must be truncated before being fed into these models, which may lead to the loss of semantically important information and potentially affect the accuracy of code smell detection.}

\begin{figure}
    \centering
    \includegraphics[width=1\linewidth]{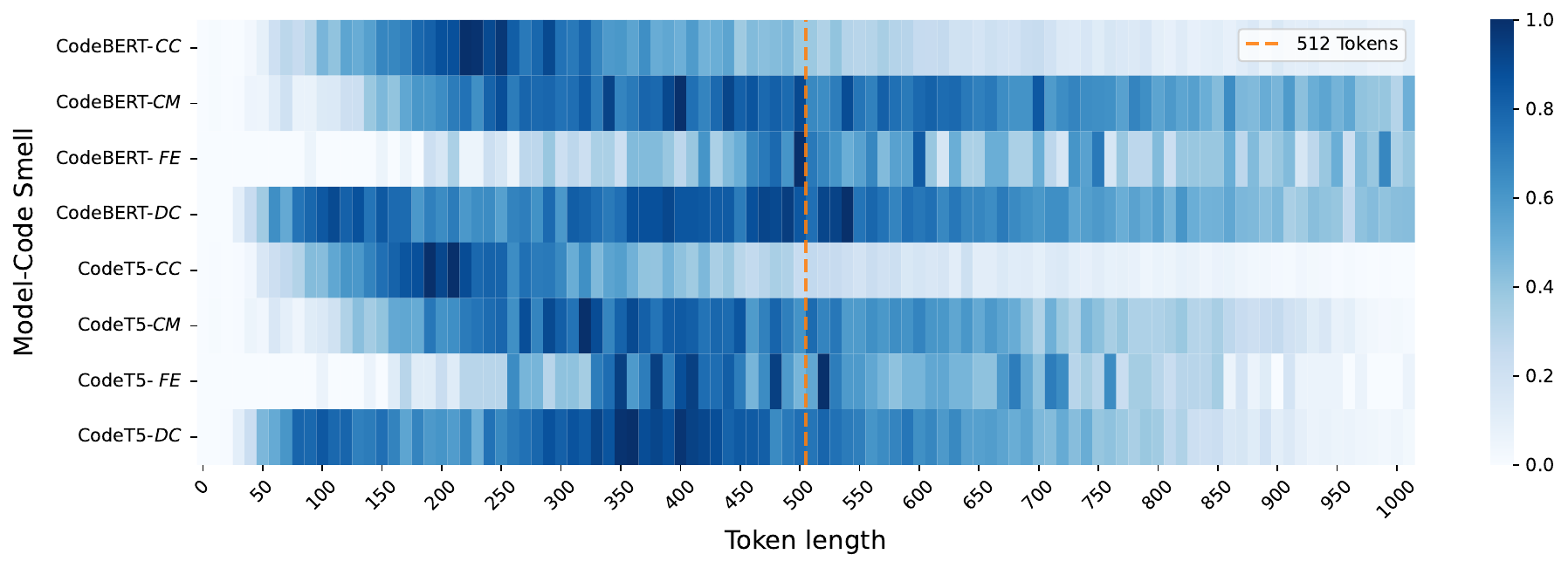}
    \caption{\textcolor{black}{Token length distribution for CodeBERT and CodeT5 across \textit{CC}, \textit{CM}, \textit{FE}, and \textit{DC}}}
    \label{fig:token length distribution}
\end{figure}

\textcolor{black}{To quantify this effect, Figure \ref{fig:token length distribution} presents the token length distributions for CodeBERT and CodeT5 across different code smell categories. GraphCodeBERT shares the same tokenizer as CodeBERT and therefore exhibits the same distribution, so only the results for CodeBERT are shown. The figure indicates that the majority of samples contain fewer than 512 tokens for all smell types. Although some samples slightly exceed the 512-token boundary, almost all remain below 600 tokens. This suggests that truncation beyond 512 tokens affects only a small portion of the dataset.}

\textcolor{black}{To further evaluate whether the truncated segments contain information that is critical for code smell detection, we conducted an additional manual analysis using CodeBERT as a representative model. Before the main review, we performed a pilot annotation: for each code smell type, we randomly selected 10 positive and 10 negative samples, which were independently labeled by the first and fourth authors. Across all smell types, the two reviewers agreed on 85-95\% of cases, indicating high consistency and supporting the reliability of our labeling.
Following this pilot, we proceeded with the main manual review.
For each code smell type, we randomly sampled 50 positive and 50 negative instances whose lengths exceed the 512-token limit. These samples were manually reviewed to determine whether the truncated portion included code elements that was essential for correct classification. Table \ref{tab:Number of truncated samples in which the truncated portion does not affect the detection of CC, CM, FE, and DC} summarizes the results.}



\begin{table}[htbp]
\textcolor{black}{
\centering
\footnotesize
\caption{\textcolor{black}{Number of truncated samples in which the truncated portion does not affect the detection of \textit{CC}, \textit{CM}, \textit{FE}, and \textit{DC}}}
\label{tab:Number of truncated samples in which the truncated portion does not affect the detection of CC, CM, FE, and DC}
\begin{tabular}{m{1.28cm}<{\centering\arraybackslash}|m{1.28cm}<{\centering\arraybackslash}|m{1.28cm}<{\centering\arraybackslash}|m{1.28cm}<{\centering\arraybackslash}|m{1.28cm}<{\centering\arraybackslash}|m{1.28cm}<{\centering\arraybackslash}|m{1.28cm}<{\centering\arraybackslash}|m{1.28cm}<{\centering\arraybackslash}}
\toprule
\multicolumn{2}{c|}{\textbf{CC}} & \multicolumn{2}{c|}{\textbf{CM}} & \multicolumn{2}{c|}{\textbf{FE}} & \multicolumn{2}{c}{\textbf{DC}} \\ \cmidrule{1-8}
\textbf{\#Positives} & \textbf{\#Negatives} & \textbf{\#Positives} & \textbf{\#Negatives} & \textbf{\#Positives} & \textbf{\#Negatives} & \textbf{\#Positives} & \textbf{\#Negatives} \\ \cmidrule{1-8}
49 & 49 & 50 & 47 & 50 & 49 & 50 & 48 \\ \cmidrule{1-8}
\end{tabular}
}
\end{table}

\textcolor{black}{The results show that for nearly all sampled instances, truncation beyond 512 tokens does not remove information critical for code smell detection. Only a few cases (1–3 samples per category) potentially lose relevant information. Most essential indicators appear within the first 512 tokens, while the truncated segments typically contain auxiliary logic or repeated structures that do not affect classification.}

\textcolor{black}{Overall, these findings, combined with the token length statistics in Figure \ref{fig:token length distribution}, suggest that context window limitations of CodeBERT, GraphCodeBERT, and CodeT5 have minimal impact on evaluation results. Nonetheless, we acknowledge that this review was conducted on a limited subset; for exceptionally long or complex methods, truncation could still obscure semantically important content. Future work could extend this analysis to a larger portion of the dataset to further quantify any residual effects more comprehensively.}

\end{document}